\newif\ifconfver
\confvertrue        

\newif\ifplainver  
\plainvertrue

\newif\ifhide  
\hidetrue

\ifplainver
    \confverfalse   
\fi

\ifconfver
     \documentclass[10pt,twocolumn,twoside]{IEEEtran}
\else
    \ifplainver
        \documentclass[11pt]{article}
        \usepackage{fullpage}
    \else
        \documentclass[12pt,draftcls,onecolumn]{IEEEtran}
    \fi
\fi

\usepackage{etoolbox}%
\usepackage{xpatch}
\usepackage{blindtext}
\usepackage{setspace}

\usepackage{tocloft}%
\newlength{\articlesectionshift}%
\let\LaTeXStandardSection\section

\let\LaTeXStandardTheSubSection\thesubsection
\let\LaTeXStandardTheSubSubSection\thesubsubsection
\let\LaTeXStandardTheParagraph\theparagraph

\makeatletter
\newcounter{titlecounter}

\xpretocmd{\maketitle}{\ifnumgreater{\value{titlecounter}}{1}}{\clearpage}{}{} 
\xpatchcmd{\maketitle}{\let\maketitle\relax\let\@maketitle\relax}{\refstepcounter{titlecounter}\begingroup

\makeatother \addtocontents{toc}{\begingroup\addtolength{\cftsecindent}{-\articlesectionshift}}%
	\addcontentsline{toc}{section}{\protect{\numberline{\thetitlecounter}{\@title~ \@author}}}%
	\addtocontents{toc}{\endgroup}
}{%
	\typeout{Patching was successful}
}{%
	\typeout{patching failed}
}%

\def\@IEEEdestroythesectionargument#1{\LaTeXStandardSection{#1}}%

\xapptocmd{\maketitle}{%
	\renewcommand{\thesection}{\LaTeXStandardTheSection}%
	\renewcommand{\thesubsection}{\LaTeXStandardTheSubSection}%
	\renewcommand{\thesubsubsection}{\LaTeXStandardTheSubSubSection}%
	\renewcommand{\theparagraph}{\LaTeXStandardTheParagraph}%
}{}{}%

\@addtoreset{section}{titlecounter}

\usepackage{calc,amsfonts,amssymb,amsmath,bm,url,color,theorem,graphicx,cite}
\usepackage{algorithm}
\usepackage{algorithmic}
\usepackage{soul}
\usepackage{enumerate}
\usepackage{bbm}
\usepackage{shortcuts_OPT}
\usepackage{multirow}
\usepackage{subcaption}
\usepackage{array}
\usepackage{diagbox}
\usepackage{pifont}%
\newcounter{mytempeqncnt}
\usepackage{dblfloatfix}

\ifconfver \else \usepackage{titling} \fi
\usepackage{eqparbox}

\definecolor{orange}{RGB}{255,107,0}
\def\blue{\color{blue}}
\def\red{\color{red}}

\newtheorem{Fact}{Fact}
\newtheorem{Lemma}{Lemma}
\newtheorem{Prop}{Proposition}

\theorembodyfont{\rmfamily}

\newcommand\bW{\ensuremath{{\bm W}}}

\newcommand\bq{\ensuremath{{\bm q}}}
\newcommand\bx{\ensuremath{{\bm x}}}
\newcommand\by{\ensuremath{{\bm y}}}

\newcommand\bh{\ensuremath{{\bm h}}}
\newcommand\bH{\ensuremath{{\bm H}}}
\newcommand\cH{\ensuremath{\bm{\mathcal{H}}}}

\newcommand\be{\ensuremath{{\bm e}}}
\newcommand\bz{\ensuremath{{\bm z}}}
\newcommand\bp{\ensuremath{{\bm p}}}
\newcommand\bP{\ensuremath{{\bm P}}}
\newcommand\bR{\ensuremath{{\bm R}}}

\newcommand\bX{\ensuremath{{\bm X}}}

\newcommand\ba{\ensuremath{{\bm a}}}

\newcommand\bA{\ensuremath{{\bm A}}}

\newcommand\bb{\ensuremath{{\bm b}}}
\newcommand\bg{\ensuremath{{\bm g}}}
\newcommand\bB{\ensuremath{{\bm B}}}

\newcommand\bmu{\ensuremath{{\bm \mu}}}
\newcommand\balp{\ensuremath{{\bm \alpha}}}
\newcommand\bbeta{\ensuremath{{\bm \beta}}}

\newcommand\bF{\ensuremath{{\bm F}}}

\newcommand\bD{\ensuremath{{\bm D}}}

\newcommand\bu{\ensuremath{{\bm u}}}
\newcommand\bnu{\ensuremath{{\bm \nu}}}
\newcommand\bv{\ensuremath{{\bm v}}}
\newcommand\bSig{\ensuremath{{\bm \Sigma}}}

\newcommand\bPhi{\ensuremath{{\bm \Phi}}}

\newcommand\btheta{\ensuremath{{\bm \theta}}}
\newcommand\tbtheta{\ensuremath{\tilde{\bm \theta}}}
\newcommand\bvartheta{\ensuremath{{\bm \vartheta}}}

\newcommand\bzeta{\ensuremath{{\bm \zeta}}}
\newcommand\bpi{\ensuremath{{\bm \pi}}}

\newcommand\bV{\ensuremath{{\bm V}}}
\newcommand\bU{\ensuremath{{\bm U}}}
\newcommand\bs{\ensuremath{{\bm s}}}

\newcommand{\Rbb}{\mathbb{R}}
\newcommand{\Cbb}{\mathbb{C}}

\newcommand{\setD}{\mathcal{D}}

\newcommand{\setX}{\mathcal{X}}

\newcommand{\setS}{\mathcal{S}}

\newcommand{\setN}{\mathcal{N}}

\newcommand{\Exp}{\mathbb{E}}
\newcommand{\jj}{\mathfrak{j}}

\newcommand{\Diag}{\mathrm{Diag}}

\newcommand\br{\ensuremath{{\bm r}}}

\newcommand{\eps}{\varepsilon}
\newcommand{\bzero}{{\bm 0}}

\newcommand{\bI}{{\bm I}}

\newcommand\bigO{\ensuremath{{\mathcal{O}}}}

\newcommand\prox{\ensuremath{{\rm prox}}}
\newcommand\Prox{\ensuremath{{\rm Prox}}}

\newcommand\dom{\ensuremath{{\rm dom}\,}}

\newcommand\her{\ensuremath{{\sf H}}}
\newcommand\Ind{\ensuremath{{\mathbbm I}}}

\newcolumntype{M}[1]{>{\centering\arraybackslash}m{#1}}

\usepackage{tikz}
\usetikzlibrary{arrows.meta}
\usetikzlibrary{decorations}
\usetikzlibrary{calc}
\usetikzlibrary{shapes.geometric}
\usetikzlibrary{external}

\begin{document}


\newcommand{\papertitle}{
Accelerated and Deep Expectation Maximization for One-Bit MIMO-OFDM Detection
}

\newcommand{\paperabstract}{
In this paper we study the expectation maximization (EM) technique for one-bit MIMO-OFDM detection (OMOD).
Arising from the recent interest in massive MIMO with one-bit analog-to-digital converters,
OMOD is a massive-scale problem.
EM is an iterative method that can exploit the OFDM structure to process the problem in a per-iteration efficient fashion.
In this study we analyze the convergence rate of EM for a class of approximate maximum-likelihood OMOD formulations,
or, in a broader sense, a class of problems involving regression from quantized data.
We show how the SNR and channel conditions can have an impact on the convergence rate.
We do so by making a connection between the EM and the proximal gradient methods in the context of OMOD.
This connection also gives us insight to build new accelerated and/or inexact EM schemes.
The accelerated scheme has faster convergence in theory, and the inexact scheme provides us with the flexibility to implement EM more efficiently, with convergence guarantee.
Furthermore we develop a deep EM algorithm, wherein we take the structure of our inexact EM algorithm and apply deep unfolding to train an efficient structured deep net.
Simulation results show that our accelerated exact/inexact EM algorithms run much faster than their standard EM counterparts,
and that the deep EM algorithm gives promising detection  and  runtime performances.
}


\ifplainver


    \title{\papertitle}

    \author{
   	Mingjie Shao$^{\dag\S}$, Wing-Kin Ma$^{\S}$, Junbin Liu$^{\S}$, and Zihao Huang$^{\S}$$^{\ddag}$\\~\\
   $^{\dag}$ School of Information Science and Engineering, Shandong University, Qingdao, China\\
$^{\S}$Department of Electronic Engineering, The Chinese University of Hong Kong, \\
    Hong Kong SAR of China\\
    $^{\ddag}$Tencent Technology (Shenzhen) Co., Ltd., Shenzhen, Guangdong, China}

    \maketitle

    \begin{abstract}
    \paperabstract
   \end{abstract}

\else
    \title{\papertitle}

    \ifconfver \else {\linespread{1.1} \rm \fi

    \author{Mingjie Shao, Wing-Kin Ma, Junbin Liu, and Zihao Huang}

    \maketitle

    \ifconfver \else
        \begin{center} \vspace*{-2\baselineskip}
        \end{center}
    \fi

    \begin{abstract}
    \paperabstract
    \end{abstract}


    \begin{IEEEkeywords}\vspace{-0.0cm}
	one-bit MIMO-OFDM detection, expectation maximization, convergence analysis, deep unfolding
    \end{IEEEkeywords}

    \ifconfver \else \IEEEpeerreviewmaketitle} \fi

 \fi

\ifconfver \else
    \ifplainver \else
        \newpage
\fi \fi


\makeatletter{\renewcommand*{\@makefnmark}{}
	\footnotetext{This work was supported by a General Research Fund (GRF) of the Research Grant Council (RGC), Hong Kong, under Project ID CUHK 14207318. The conference version of this paper appeared in ICASSP 2021.}\makeatother}


\section{Introduction}

In massive multiple-input multiple-output (MIMO),
we have recently seen enormous interest in techniques that are related to low-resolution or coarsely quantized (CQ) signals from a fundamental signal processing viewpoint.
This trend is driven by our interest in replacing high-resolution analog-to-digital converters (ADCs) and digital-to-analog converters (DACs) with lower resolution ones,
so that we can reduce the hardware cost and energy consumption of the radio-frequency (RF) front ends.
In the conventional scenario of high-resolution ADCs/DACs, quantization errors are insignificant and are typically treated as part of the background noise.
In the low-resolution scenario, recent research has suggested that we should exploit the low-resolution or quantization structure to better harness the problem.
For example, for MIMO downlink precoding with one-bit DACs,
a popular direction is to directly design the one-bit MIMO transmit signals by discrete optimization;
see, e.g., \cite{Jacobsson2017,Jedda2018,Sohrabi2018,shao2019framework} and the references therein.
In addition to exploiting the low-resolution structure, there have been endeavors that apply Sigma-Delta modulation---a widely-used technique in temporal DACs/ADCs---to spatially shape the quantization errors such that users in a designated angular sector will
experience
much less quantization error effects \cite{shao2019one,rao2021massive}.

Another meaningful problem in massive MIMO with low-resolution ADCs/DACs is MIMO detection.
The problem typically appears in the uplink, and the goal is to detect signals transmitted by different users over the same RF band.
While numerous MIMO detection methods were already developed in the conventional high-resolution ADC scenario,
they do not assume the presence of quantization and may not be directly applied to the low-resolution ADC scenario.
We have indeed seen research on applying linear receivers to the low-resolution scenario \cite{mezghani2007modified,risi2014massive,li2017channel},
but recent research appears to favor exploiting the quantization structure.
The key idea that underlies the majority of low-resolution or CQ MIMO detection methods is that we consider maximum likelihood (ML) or related inference approaches, and the likelihood function is derived from quantized signals (quantization is treated as a function, and we do not model quantization errors as noise).
This rationale gave rise to a variety of CQ MIMO detection methods;
e.g., expectation maximization (EM) 
\cite{plabst2018efficient},
convex and non-convex proximal gradient (PG) methods \cite{choi2016near,studer2016quantized,shao2020binary},
deep learning
\cite{shao2020binary,nguyen2021linear,khobahi2021lord},
and posterior inference via approximate message passing \cite{wen2015bayes} and variational Bayes \cite{thoota2022massive}.

\subsection{The Problem of Interest: CQ MIMO-OFDM Detection}

In MIMO detection it is common to assume frequency-flat channels.
Yet, in the current mobile systems, we consider frequency-selective channels.
This is not a problem in the high-resolution scenario,
because the current systems usually use the orthogonal frequency division multiplexing (OFDM) scheme,
and OFDM breaks the frequency-selective channels into many parallel frequency-flat channels.
This means that the MIMO detection methods for frequency-flat channels can be directly applied.
Unfortunately, the same does not hold for the CQ case.
The presence of quantization destroys the structure required to turn the problem into parallel frequency-flat channels.
Subsequently, when we consider MIMO-OFDM detection with low-resolution ADCs, we have to engage with the full form of the problem.
The arising difficulty  is that the problem is a massive-scale problem;
the problem size scales with the OFDM size, which is of the order of hundreds.
While in principle we can directly apply the CQ MIMO detection methods in the frequency-flat channel case to the MIMO-OFDM case,
it is computationally infeasible to do so due to the massive problem size.
It is necessary to exploit the OFDM structure to efficiently realize MIMO detection methods.

There are far fewer studies for CQ MIMO-OFDM detection~\cite{Mirfarshbafan2020,plabst2018efficient,garcia2021channel,thoota2022massive}.
In \cite{Mirfarshbafan2020}, the PG method for the box ML relaxation formulation was developed.
The PG method is able to exploit the OFDM structure to have per-iteration efficient implementation.
In \cite{plabst2018efficient,garcia2021channel}, the EM method for the Gaussian maximum a posteriori formulation  was built.
The EM method is iterative,
and at each iteration we solve a problem---the so-called M-step problem---that takes the same form as MIMO-OFDM {\em without} quantization.
Hence we can exploit the OFDM structure to have per-iteration efficient implementation,
and this makes the EM method appealing in OMOD.
The same benefit was also seen in the application of the variational Bayes method (which is related to EM) to CQ MIMO-OFDM detection \cite{thoota2022massive}.

\subsection{This Work}

In this paper we study the EM technique for CQ MIMO-OFDM detection.
In particular, we consider MIMO-OFDM detection with one-bit ADCs, or simply one-bit MIMO-OFDM detection (OMOD).
We are interested in
 EM convergence rate analysis for a class of approximate ML OMOD formulations.
This aspect has not been seriously studied in CQ MIMO detection. 
Our analysis draws insights from the analysis techniques in  PG methods \cite{nesterov2007gradient,beck2009fast,schmidt2011convergence,beck2017first}.
We will reveal a  hidden  connection between the PG and EM methods,
and from that connection we establish EM convergence results by taking key ideas from the PG analysis techniques.
We will show that
(a) EM converges at a rate of at least $1/k$, where $k$ denotes the iteration number;
that
(b) EM should converge faster than PG, in theory.
Also, we show how the SNR and channel conditions can have an impact on the convergence rate.
In particular, our analysis indicates that the convergence rate is slower as the SNR is higher; this applies to both PG and EM.
This high-SNR slow convergence phenomenon was noticed by numerical results in previous studies \cite{shao2020binary,Mirfarshbafan2020},
and, to the best of our knowledge, this is the first time we see a theoretical result for such phenomenon.

Our EM convergence study also leads to new OMOD algorithms.
In first-order optimization, it is well-known that the PG method can be accelerated to a convergence rate of at least $1/k^2$ if we employ extrapolation \cite{nesterov2007gradient,beck2009fast}; we assume convex problems.
Using the connection between the PG and EM methods,
we apply the acceleration idea to EM.
The result is an accelerated EM scheme for OMOD, which has the $1/k^2$ convergence rate guarantee and runs faster than the standard EM as will be shown in our numerical results.
In addition we consider an inexact EM scheme,
where we want to solve the M-step problems not too precisely to save complexity.
By taking insight from the study of inexact PG \cite{schmidt2011convergence}, we provide theoretical conditions under which the inexact EM scheme will have convergence guarantees.
The inexact EM scheme can be used together with acceleration, and our numerical results will demonstrate that the accelerated inexact EM algorithm for OMOD has significant merits in runtime savings.

Lately we have seen much interest in applying deep learning to MIMO detection \cite{Samuel2019learning,corlay2018multilevel,He2020model,shao2020binary,khobahi2021lord,shlezinger2020deepsic}.
The most successful approach at present is deep unfolding \cite{gregor2010learning}. 
The principle is to see a model-based algorithm as a deep net, take the structure, and use data-driven learning to try to learn a better deep net.
We apply deep unfolding to our inexact EM algorithm for OMOD;
that is, simply speaking, we take the structure of how efficient EM can handle MIMO-OFDM, and we use that to build an efficient highly-structured deep net.
Our numerical results will show that the deep inexact EM algorithm is competitive in both detection and runtime performances, compared to the previous OMOD algorithms.

It is worth noting that the convergence analysis and the accelerated and inexact EM schemes in this study are, in a broad sense, also applicable to a variety of formulations arising from regression from CQ data,
such as CQ MIMO channel estimation \cite{ivrlac2007mimo,mezghani2010multiple} and CQ compressive sensing \cite{zymnis2009compressed}.

\subsection{Related Works}

We should discuss related works.
Our EM convergence analysis plays a large part in this study,
and one may question if the existing convergence analyses \cite{wu1983convergence,razaviyayn2013unified,mairal2013optimization,hong2017iteration} provide the same or better results.
The existing studies mostly consider a general framework, covering a variety of optimization schemes and a rich class of problems.
Our study, on the other hand, exploits the specific structure of our OMOD problem; e.g., the previously mentioned result of high-SNR slow convergence is unique to our problem.
In terms of the analysis techniques used,
our analysis is considered closest to Mairal's study \cite{mairal2013optimization}, which considers a general majorization minimization (MM) framework and can cover EM.
Mairal's study takes insight from the PG analysis techniques
through some assumptions.
We also draw insight from the PG analysis techniques, but we do so by best using our problem's specific structure.
As we will discuss in details, our result arguably provides better characterization of the convergence  behaviors.

One key result in this study is accelerated EM for OMOD.
Indeed, Mairal's study \cite{mairal2013optimization} has  described the same kind of acceleration for MM.
To the best of our knowledge, no application was provided in Mairal's study and in subsequent studies.
Our application is probably the first, showing the efficiency of such accelerated EM or MM.
We should also discuss the application of deep unfolding.
While deep unfolding has been widely considered in MIMO detection,
it has not been used in CQ MIMO-OFDM detection.

The preliminary version of this study was presented in a conference \cite{shao2021divide}.
It studied inexact EM as a working idea.
It does not consider EM convergence analysis, accelerated EM and deep EM, which are the main contributions in this paper.

\subsection{Organization and Notations}

This paper is organized as follows.
Section~\ref{sect:em_basics}, \ref{sect:OMOD} and \ref{sect:OMOD_SOTA} review the EM basics, the MIMO-OFDM detection problem and the OMOD problem, respectively.
In Section~\ref{sect:conv}, we perform EM convergence analysis and establish the accelerated and inexact EM methods.
Section~\ref{sect:AIEM} develops an accelerated inexact EM algorithms for OMOD.
Section~\ref{sect:DIEM} builds a deep inexact EM algorithm via deep unfolding.
Section~\ref{sec:sim} provides the numerical results, and we conclude in Section~\ref{sect:conclusion}.

Some basic notations are described as follows.
The sets of real, non-negative, non-positive and complex numbers are denoted by $\Rbb$, $\Rbb_+$, $\Rbb_-$ and $\Cbb$, respectively;
we denote $\jj = \sqrt{-1}$;
boldface lowercase letters, e.g., $\bx$, represent column vectors;
the $i$th element of a vector $\bx$ is denoted by $x_i$;
boldface capital letters, e.g., $\bX$, represent matrices;
the superscripts $^\top$, $^\her$, $^{-1}$, and $^\dag$ denote the transpose, Hermitian transpose, inverse, and pseudoinverse, respectively;
$\Re(\bx)$ and $\Im(\bx)$ denote the real and imaginary parts of $\bx$, respectively;
$\bx = (\bx_1,\ldots,\bx_m)$ means that
$\bx = [~ \bx_1^\top, \ldots, \bx_m^\top ~]^\top$;
$\| \cdot \|$ denotes the Euclidean norm;
$\bzero$ and $\bI$ denote an all-zero vector and an identity matrix, respectively;
$\Diag(\bx)$ denotes a diagonal matrix whose $(i,i)$th element is $x_i$;	
$\bx \sim \setD$ means that $\bx$ is a random vector following a probability distribution $\setD$;
$\setN(\bmu, \bSig)$ denotes a multivariate Gaussian distribution with mean $\bmu$ and covariance $\bSig$; $\setN(\bx; \bmu, \bSig)$ denotes the distribution function corresponding to $\setN(\bmu, \bSig)$.

\section{A Review of EM for Quantized Regression}
\label{sect:em_basics}

Before we delve into the details of OMOD, we would like to first review the classic concepts of expectation maximization (EM) \cite{dempster1977maximum} in the context of quantized linear regression (QLR).
In particular, the arising notion of iterative linear regression is worth appreciating, and we will need some of the details later.

We consider a basic QLR problem for which we have an observation $\by \in \{ \pm 1 \}^m$ following a model
\beq \label{eq:exa_1bit_model}
\by = {\rm sgn}(\bx), \quad \bx = \bA \btheta + \bv,
\eeq
where
${\rm sgn}: \Rbb^m \rightarrow \{ \pm 1 \}^m$ is the sign function;
$\bA \in \Rbb^{m \times n}$ is known;
$\btheta \in \Rbb^n$ is a model parameter;
$\bv \sim \setN(\bzero,\sigma^2 \bI)$ is noise;
$\sigma^2$ is known.
Note that $\bx$ is unobserved.
The problem is to estimate $\btheta$ from $\by$.
As we will see later, OMOD can be viewed as an instance of QLR, with $\bA$ and $\btheta$ having certain structures.
We consider the maximum-likelihood (ML) estimation
\beq \label{eq:exa_ml}
\hat{\btheta} \in \arg \max_{\btheta \in \Rbb^n} \, L(\btheta) := \log p(\by|\btheta),
\eeq
where $p(\by|\btheta)$ is the distribution of $\by$ given $\btheta$; the same notation will be used to denote the distributions of other random variables.
It can be shown that
$p(\by|\btheta) = \prod_{i=1}^m p(y_i|\btheta)$,
\[
p(y_i|\btheta)  = \int_{\setX(y_i)} p(x_i|\btheta) {\rm d}x_i,
\quad
p(x_i|\btheta) = \setN( x_i; \ba_i^\top \btheta, \sigma^2 ),
\]
where
$\ba_i$ denotes the $i$th row of $\bA$; $\setX(y_i) = \Rbb_+$ if $y=1$, and $\setX(y_i) = \Rbb_-$ if $y=-1$.
This leads to
\beq \label{eq:exa_Like}
L(\btheta) =
\sum_{i=1}^m \log \Phi \left( \frac{ y_i \ba_i^\top \btheta }{\sigma}  \right),
\eeq
where
$\Phi(z) = \int_{-\infty}^z e^{- t^2/2}  {\rm d}t / \sqrt{2\pi}$ is the standard cumulative Gaussian distribution.
It is known that
$L$ is concave \cite{boyd2004convex,choi2016near}.

The ML estimation problem \eqref{eq:exa_ml} has no closed-form solution in general, and EM is a popular way to solve the ML problem.
Intuitively, the idea is to use
the fact that estimating $\btheta$ is easy if $\bx$ is observable.
To describe the EM method, let
\[
G(\btheta|\btheta')  = \underbrace{\Exp_{\bx} [ \log( p(\bx|\btheta )) | \by, \btheta' ]}_{:= Q(\btheta|\btheta')} -   \underbrace{\Exp_{\bx} [ \log( p(\bx|\by,\btheta' )) | \by, \btheta' ]}_{:= H(\btheta')}.
\]
Here, $\Exp_\bx[ f(\bx) | \by, \btheta' ] = \int_\setX f(\bx) p(\bx|\by,\btheta') {\rm d}\bx$ denotes a conditional expectation
of
$\bx$; $\setX$ denotes the support of $p(\bx|\by,\btheta')$.
It is known that
$G(\btheta|\btheta') \leq L(\btheta)$ for all $\btheta, \btheta'$, and $G(\btheta'|\btheta')= L(\btheta')$ for all $\btheta'$;
i.e., $G(\btheta|\btheta')$ is a lower bound of $L(\btheta)$.
EM considers a successive lower-bound approximation
\beq \label{eq:exa_EMM}
\btheta^{k+1} \in \arg \max_{\btheta \in \Rbb^n} G(\btheta|\btheta^k), ~ k=0, 1, 2,\ldots
\eeq
An important question is whether $Q(\btheta|\btheta')$ is tractable.
By noting $p(\bx|\btheta) = \setN(\bx;\bA\btheta, \sigma^2 \bI)$, one can verify that
\beq \label{eq:Qtt}
\begin{aligned}
Q(\btheta|\btheta')
& = -\tfrac{1}{2\sigma^2} \| \Exp_{\bx}[ \bx | \by, \btheta' ] - \bA\btheta \|^2 + {\rm const.},
\end{aligned}
\eeq
where ``const.'' does not depend on $\btheta$.
It can be shown that
\beq \label{eq:exa_EMM3}
\Exp_{x_i}[ x_i | \by, \btheta' ]
= \ba_i^\top \btheta' + y_i \sigma \frac{\phi( y_i \ba_i^\top \btheta' /\sigma )}{\Phi( y_i \ba_i^\top \btheta' /\sigma)},
\eeq
where $\phi(t) =  e^{- t^2/2} / \sqrt{2\pi}$;
see, e.g., \cite{ivrlac2007mimo}, for details.
To prove \eqref{eq:exa_EMM3}, one needs to show
that
$p(x_i|y_i,\btheta)$
is a truncated Gaussian distribution,
namely, $p(x_i|y_i,\btheta) = p(x_i|\btheta)/p(y_i|\btheta)$, with $\setX(y_i)$ being the support;
that $\Exp_{x_i}[ x_i | \by, \btheta' ]$ is the mean of $p(x_i|y_i,\btheta')$;
that
a truncated Gaussian mean has an explicit expression \cite{ivrlac2007mimo}.
The EM method can hence be written as
\beq \label{eq:exa_EMM1}
\btheta^{k+1} \in \arg \min_{\btheta \in \Rbb^n} \| \bx^k - \bA \btheta \|^2, ~ k=0, 1, 2,\ldots,
\eeq
where $\bx^k = \Exp_{\bx}[ \bx | \by, \btheta^k ]$ is computed by \eqref{eq:exa_EMM3}.
The result is an {\em iterative linear regression}:
we estimate $\bx$ by the conditional mean $\bx^k$, inferred from $\by$ and the previous estimate $\btheta^k$;
and then we estimate $\btheta$ by the linear regression in \eqref{eq:exa_EMM1},
which is easy to solve.
This iterative linear regression process looks appealing and makes sense.


Some history should be noted.
The EM method  for
 QLR
 dates back to as early as 1980  in statistics, with application to blood lead data \cite{hasselblad1980analysis}.
It also arose in signal processing, with applications to MIMO channel estimation and detection from coarsely quantized  signals \cite{ivrlac2007mimo,mezghani2010multiple,plabst2018efficient,garcia2021channel}.

\section{A Review of the MIMO-OFDM Background}
\label{sect:OMOD}

In this section, we review some background of MIMO-OFDM detection.
Section~\ref{sect:omod_model} describes the MIMO-OFDM signal model.
Section~\ref{sect:mimo_ofdm_uq_2} shows the subcarrier decoupling concept, which, as will be seen later, plays a pivotal role in the application of EM to OMOD.


\subsection{MIMO-OFDM Model}
\label{sect:omod_model}

Consider a wireless uplink scenario where a number of users simultaneously send signals to a multi-antenna base station (BS).
The transmitted signals undergo frequency selective channel effects, upon reception at the BS; and the user side employs the OFDM transmission scheme.
The received signals at the BS can be modeled as
\beq \label{eq:model_uq}
\br_m = \textstyle \sum_{n=1}^N \bH_{m,n} \bF^\her \bs_n + \bnu_m, \quad m=1,\ldots,M,
\eeq
where $\br_m \in \Cbb^W$ is a block of discrete-time signals received by the BS's $m$th antenna;
$W$ is the block length, or the OFDM size;
$M$ is the number of antennas at the BS;
$\bF \in \Cbb^{W \times W}$ is the unitary discrete Fourier transform (DFT) matrix;
$\bs_n \in \Cbb^W$ is a block of symbols sent by user $n$;
$N$ is the number of users;
$\bH_{m,n} \in \Cbb^{W \times W}$ is a circulant matrix, constructed from a discrete-time impulse response $\bh_{m,n} = (h_{0,m,n}, \ldots, h_{W'-1,m,n}, 0, \ldots, 0) \in \Cbb^W$ of the channel from user $n$ to the BS's $m$th antenna, with $W' \ll W$;
$\bnu_m$ is circular complex Gaussian noise with mean $\bzero$ and covariance $\sigma_{\rm C}^2 \bI$, and with $\bnu_1,\ldots,\bnu_M$ being independent.
The symbols $s_{w,n}$'s are drawn from a QAM constellation set
\beq \nonumber \\
\begin{aligned}
	\setS & = \{ s \in \Cbb | \mid \Re(s), \Im(s) \in \{ \pm 1, \pm 3, \ldots, \pm (2D-1) \} \},
\end{aligned}
\eeq
where $D$ is a positive integer.
The above model
is standard, and
we refer the reader to the literature (e.g., \cite{goldsmith2005wireless}) for details.
For conciseness, we let $\br = (\br_1,\ldots,\br_M)$ and express \eqref{eq:model_uq} as
\beq \label{eq:model_uq_2}
\br = \cH \bs + \bnu,
\eeq
where $\bs = (\bs_1,\ldots,\bs_N)$,
$\bnu = (\bnu_1,\ldots,\bnu_M)$,
\beq \label{eq:cH}
\cH= \begin{bmatrix}
	\bH_{1,1} \bF^\her & \hdots & \bH_{1,N} \bF^\her \\
	\vdots & & \vdots \\
	\bH_{M,1} \bF^\her & \hdots & \bH_{M,N} \bF^\her
\end{bmatrix}.
\eeq


\subsection{Subcarrier Decoupling in Unquantized MIMO-OFDM}
\label{sect:mimo_ofdm_uq_2}

Consider the problem of ML detection of the symbol vector $\bs$ from the received signal $\br$, given information of the channel $\cH$.
It is well known that the ML detection problem can be formulated as
\beq \label{eq:ml_uq}
\hat{\bs} \in \arg \min_{\bs \in \setS^{NW}}
\| \br - \cH \bs  \|^2.
\eeq
At first look, this is a massive-scale problem;
the OFDM size $W$ is often of the order of hundreds.
But it is widely known that \eqref{eq:ml_uq} can be decoupled into many smaller problems.
Since $\bH_{m,n}$ is circulant, we have the eigendecomposition
\beq \label{eq:H_evd}
\bH_{m,n}= \bF^\her \bD_{m,n} \bF, \quad \bD_{m,n} = \Diag(\tilde{\bh}_{m,n}),
\eeq
where $\tilde{\bh}_{m,n} = \bF \bh_{m,n}$.
Let $\tilde{\br}_m = \bF \br_m$. One can show that
\beq \label{eq:sub_dec} \nonumber
\begin{aligned}
	\| \br - \cH \bs  \|^2
	&  = \textstyle \sum_{m=1}^M \| \tilde{\br}_m - \sum_{n=1}^N \bD_{m,n}\bs_n \|^2 \\
	&  = \textstyle \sum_{w=1}^W \| \check{\br}_w - \check{\bH}_w \check{\bs}_w \|^2,
\end{aligned}
\eeq
where $\check{\br}_w= (\tilde{r}_{1,w},\ldots,\tilde{r}_{M,w}) \in \Cbb^M$, $\check{\bH}_w = [ \tilde{h}_{w,m,n} ]_{m,n} \in \Cbb^{M \times N}$
and $\check{\bs}_w= ({s}_{w,1},\ldots,{s}_{w,N}) \in \Cbb^N$ are the received signal, MIMO channel response and symbol vector at subcarrier $w$, respectively.
We can hence rewrite \eqref{eq:ml_uq} as
\beq \label{eq:ml_uq_subdec}
\min_{\check{\bs}_w \in \setS^N} \| \check{\br}_w - \check{\bH}_w \check{\bs}_w \|^2,
\text{~ for $w=1,\ldots,W$.}
\eeq
We see that each problem in \eqref{eq:ml_uq_subdec} has a much smaller size, and can be dealt with independently.

The decomposition of \eqref{eq:ml_uq} into \eqref{eq:ml_uq_subdec} will be called {\em subcarrier decoupling,} in what follows.
In MIMO-OFDM detection, subcarrier decoupling is commonly used to break the problem into many parallel smaller-size MIMO detection problems.


\section{ OMOD Problem Statement and  Prior Study}
\label{sect:OMOD_SOTA}

In this section, we describe the background of OMOD.
Sections~\ref{sect:1b_mimo_ofdm} and \ref{sect:prob_form} provide the problem formulation and statement.
Sections~\ref{sect:1c_mimo_ofdm} and \ref{sect:1d_mimo_ofdm} review the existing methods by proximal gradient and EM, respectively.



\subsection{One-Bit MIMO-OFDM Detection (OMOD)}
\label{sect:1b_mimo_ofdm}

OMOD considers the same MIMO-OFDM scenario in Section~\ref{sect:omod_model}, but with one-bit ADCs.
In the MIMO-OFDM detection in Section~\ref{sect:OMOD}, we assume that signal acquisitions at the BS have negligible quantization errors;
to put it another way, the BS is equipped with high-resolution ADCs.
For the one-bit ADC case, the acquired signals are
\beq \label{eq:ob_quant}
\bq_m = {\rm sgn}( \Re(\br_m)) + \jj \cdot  {\rm sgn}( \Im(\br_m)),
\quad m=1,\ldots,M,
\eeq
where the $\br_m$'s are the unquantized MIMO-OFDM signals in \eqref{eq:model_uq}.
As mentioned in the Introduction, the one-bit ADC case is motivated by the massive MIMO scenario where we want to use cheap ADCs to  reduce the hardware cost and energy consumption of the RF front ends.

The OMOD problem is formulated as follows.
Applying the signal model \eqref{eq:model_uq_2} to  \eqref{eq:ob_quant} gives
\begin{align}
	\label{eq:model_q_1}
	\bx & :=
	\begin{bmatrix} \Re(\br) \\ \Im(\br) \end{bmatrix}
	= \underbrace{\begin{bmatrix} \Re(\cH) & -\Im(\cH) \\ \Im(\cH) & \Re(\cH) \end{bmatrix}}_{:= \bA}
	\underbrace{\begin{bmatrix} \Re(\bs) \\ \Im(\bs) \end{bmatrix}}_{:= \btheta}
	+
	\underbrace{\begin{bmatrix} \Re(\bnu) \\ \Im(\bnu) \end{bmatrix}}_{:= \bv}, \\
	\label{eq:model_q_2}
	\by & := \begin{bmatrix}
		\Re(\bq) \\ \Im(\bq)
	\end{bmatrix}
	= {\rm sgn}(\bx),
\end{align}
where $\bq= (\bq_1,\ldots,\bq_M)$, $\bv \sim \setN(\bzero, \sigma^2 \bI)$, $\sigma^2 = \sigma_{\rm C}^2/2$.
The OMOD signal model in \eqref{eq:model_q_1}--\eqref{eq:model_q_2} is seen to take the form of the basic QLR model in \eqref{eq:exa_1bit_model} in Section~\ref{sect:em_basics},
and hence we can apply the results in Section~\ref{sect:em_basics} to OMOD.
Specifically, the ML detection of the symbol vector $\btheta$ from the one-bit MIMO-OFDM signal $\by$, given information of $\cH$ and  $\sigma^2$, can be formulated as
\beq \label{eq:ml_q}
\begin{aligned}
	\min_{\btheta \in \Rbb^{2NW}}
	& ~
	f(\btheta) := -
	\textstyle 	
	\sum_{i=1}^{2MW} \log \Phi
	( \bb_i^\top \btheta )
	\\
	{\rm s.t.} & ~ \theta_i \in \setD := \{ \pm 1, \pm 3, \ldots, \pm (2D-1) \}, ~
	\text{$\forall ~ i$,}
\end{aligned}
\eeq
where $\bb_i = y_i \ba_i/\sigma$; $\ba_i$ is the $i$th row of $\bA$; or,
\beq \label{eq:ana_B}
\bB := [~ \bb_1,\ldots,\bb_{2MW} ~]^\top =  \tfrac{1}{\sigma} \Diag(\by) \bA.
\eeq

There is an important aspect to note.
The subcarrier decoupling in Section~\ref{sect:mimo_ofdm_uq_2}, which allows us to decouple the massive-scale unquantized MIMO-OFDM problem into many smaller problems, does not work for OMOD.
This is because $f$ in \eqref{eq:ml_q} does not possess the structure required to apply subcarrier decoupling.
This means that we will have to confront the massive-scale nature of MIMO-OFDM in the one-bit case.


\subsection{Problem Statement}
\label{sect:prob_form}

Our study will revolve around the following formulation
\beq \label{eq:main_prob}
\min_{\btheta \in \Rbb^{2NW}} f(\btheta) + h(\btheta),
\eeq
 where $h:\Rbb^{n} \rightarrow \Rbb \cup \{ + \infty \}$ is a penalty function that takes a component-wise separable form
\beq
h(\btheta) = \textstyle \sum_{i=1}^{2NW} h(\theta_i).
\eeq
The above formulation is an approximation of the ML OMOD in \eqref{eq:ml_q}.
Consider the following two examples.

 \subsubsection*{1) Gaussian maximum a posteriori (GMAP) \cite{plabst2018efficient}}
	Consider
	\beq \label{eq:h_GMAP}
	h(\btheta) = \frac{\lambda}{{2}} \| \btheta \|^2,
	\eeq
	where we choose  $\lambda^{-1} = \Exp[ |\theta|^2 ] = ((2D)^2-1)/3$, assuming $\theta$ being uniformly distributed on $\setD$.
	The corresponding formulation \eqref{eq:main_prob} is maximum a posteriori estimation under a Gaussian prior; i.e., $\max_\btheta \log( p(\by|\btheta) p(\btheta))$ with prior $p(\btheta) = \setN(\btheta; \bzero, \lambda^{-1} \bI)$.
	This means that we approximate the symbols as Gaussian random variables.

 \subsubsection*{2) Box  \cite{Mirfarshbafan2020}}
	Consider
	\beq \label{eq:h_box}
	 h(\btheta) = \Ind_{[-U,U]^{2NW}}(\btheta), \quad U= 2D-1,
	\eeq
	where $\Ind_\setX(\bx) = 0$ if $\bx \in \setX$, and $\Ind_\setX(\bx) = \infty$ if $\bx \notin \setX$.
	This corresponds to a relaxation of the ML OMOD detector \eqref{eq:ml_q} by replacing the constellation set $\setD$ with an interval $[-U,U]$.
	
%
%
%

We are interested in studying
how problem \eqref{eq:main_prob}, which has a massive scale, can be handled in an efficient fashion.
The next two subsections describe two existing methods for \eqref{eq:main_prob}.


\subsection{Existing Method: Proximal Gradient}
\label{sect:1c_mimo_ofdm}

One way to handle problem \eqref{eq:main_prob} is to apply the  proximal gradient (PG) method \cite{beck2017first}.
The PG method for \eqref{eq:main_prob} is given by
\beq \label{eq:prox_grad}
\btheta^{k+1} = \prox_{\eta_k h}( \btheta^k - \eta_k \nabla f(\btheta^k) ),
\quad k=0,1,2,\ldots,
\eeq
where $\eta_k > 0$ is a step size;
$\nabla f(\btheta)$ is the gradient of $f$ at $\btheta$,
and it can be shown that
\beq \label{eq:exa_grad}
\nabla f(\btheta) =
- \sum_{i=1}^{2MW} \frac{\phi( \bb_i^\top \btheta) }{\Phi(\bb_i^\top \btheta)}  \bb_i = - \bB^\top \psi(\bB \btheta),
\eeq
for which $\psi(\bz) = (\psi(z_1),\ldots,\psi(z_m))$,
$\psi(z) = \phi(z)/\Phi(z)$;
\[
\prox_h(\bx) \in \arg \min_{\bz \in \Rbb^n} \tfrac{1}{2} \| \bx - \bz \|^2 + h(\bz)
\]
is the proximal operator associated with $h$.
We typically consider cases for which $\prox_h(\bx)$ is easy to compute.
For example, for the box case $h= \Ind_{[-U,U]^n}$, we have
\beq \label{eq:clip}
[ \prox_{h}(\bx) ]_i = \left\{ \begin{array}{ll}
	U, & x_i \geq U \\
	x_i, & x_i \in [-U,U] \\
	-U, & x_i \leq -U
\end{array} \right.
\eeq

We want to examine the complexity of the PG method.
The most expensive part lies in computing the gradients $\nabla f(\btheta^k)$.
At first sight, computing $\nabla f(\btheta)$ in \eqref{eq:exa_grad} seems expensive since $\bB$ is very large.
But one can use the OFDM structure to reduce the complexity.
It was shown in \cite{Mirfarshbafan2020} that
$\nabla f(\btheta)$ can be computed with a complexity of
\beq \label{eq:pg_complex}
\bigO( 2MW (\log(W) + N + C ) ),
\eeq
where $C$ denotes the complexity to numerically compute $\Phi$, given a precision\footnote{The function $\Phi$ has no closed form, and it is typically evaluated by an off-the-shelf algorithm (e.g., \texttt{erf} on MATLAB).}.
For the reader's convenience, in the supplemental material we describe the details with the computation of $\nabla f(\btheta)$.
Simply speaking, to compute $\nabla f(\btheta)$ efficiently, we need to run
fast Fourier transforms (FFTs) or inverse FFTs (IFFTs)
for a total of $2M$ times.
To sum up,
the per-iteration complexity of the PG method is \eqref{eq:pg_complex}.

The PG method for OMOD was studied in \cite{Mirfarshbafan2020}, wherein the efficient FFT/IFFT-based method for computing the gradients was proposed, and the box relaxation was considered.


\subsection{ Existing Method: Expectation Maximization}
\label{sect:1d_mimo_ofdm}

Another way to handle problem \eqref{eq:main_prob} is to apply the EM method.
 By following the EM method reviewed in Section~\ref{sect:em_basics}, we can show the following.
In accordance with the E-step \eqref{eq:Qtt} of the EM method in Section~\ref{sect:em_basics}, the EM surrogate function of the objective function $f$ of the OMOD problem \eqref{eq:main_prob} is given by
\begin{align}
g(\btheta|\btheta')  & = \tfrac{1}{2\sigma^2} \| {\bx}' - \bA \btheta \|^2 + {\rm const.} \nonumber \\
& = \tfrac{1}{2\sigma^2} \| {\br}' - \cH\bs\|^2 + {\rm const.} := g(\bs|\bs'),
\end{align}
where $\bx' = \Exp_{\bx}[ \bx | \by, \btheta' ]$ is a conditional mean of the unquantized signal $\bx$ in \eqref{eq:model_q_1}--\eqref{eq:model_q_2}, and, similarly, ${\br}'= \Exp_{\br}[ \br | \bq, \bs' ]$ is a conditional mean of the unquantized MIMO-OFDM signal $\br$ in \eqref{eq:model_uq_2}.
The EM method is given by
\beq \label{eq:em1}
\bs^{k+1} \in \arg \min_{\bs \in \Cbb^{NW}}
g(\bs|\bs^k)
+ h(\bs),
\eeq
where, for convenience, we denote $h(\bs) = h(\Re(\bs)) + h(\Im(\bs))$.
The merit of the EM method is that it allows us to apply  the subcarrier decoupling in Section~\ref{sect:mimo_ofdm_uq_2}.
To describe this, let
\beq \label{eq:em1_r}
\br^k_m = \Exp_{\br_m}[ \br_m | \bq, \bs^k], ~ \tilde{\br}_m^k = \bF \br_m^k, ~ \check{\br}_w^k = ( \tilde{r}_{1,w}^k,\ldots,\tilde{r}_{M,w}^k).
\eeq
By the subcarrier decoupling concept in Section~\ref{sect:mimo_ofdm_uq_2},
we can decompose \eqref{eq:em1} into
\beq \label{eq:em1_dec}
\check{\bs}_w^{k+1} \in \arg \min_{\check{\bs}_w \in \Cbb^{N}} \tfrac{1}{2\sigma^2} \| \check{\br}_w^k - \check{\bH}_w \check{\bs}_w \|^2 + h(\check{\bs}_w),
\eeq
for $w=1,\ldots,W$.
The basic description of the EM method is complete.

Intuitively, the EM method seems more promising than the PG method in the previous subsection.
The EM method has no parameter, such as step size, to select;
it inherits the merits of subcarrier decoupling, which seems to suggest better efficiency;
it performs more optimization in one iteration, specifically, \eqref{eq:em1_dec}, which seems to suggest faster convergence.

The story is actually more complex, as one tries to unravel.
We should not forget the fact that the conditional means $\br^k_m$'s need to be computed.
Examining the equations closely (see \eqref{eq:exa_EMM3}, \eqref{eq:model_uq}, \eqref{eq:model_uq_2}, and \eqref{eq:model_q_1}),
one will find that $\br_m^k$ is given by
\begin{subequations} \label{eq:condmean_comp}
	\begin{align}
		\br_m^k & = \bz_m^k + \bzeta_m^k, \\
		\bz_m^k & = \textstyle \sum_{n=1}^N \bH_{m,n}\bF^\her \bs_n^k  = \bF^\her(  \sum_{n=1}^N \bD_{m,n} \bs_n^k), \label{eq:condmean_comp:b} \\
		\Re( \bzeta_m^k )
		& = \sigma \left( \Re(\bq_m) \odot \psi\left( \tfrac{1}{\sigma} \Re(\bq_m) \odot \Re(\bz_m^k) \right)  \right), \\
		\Im( \bzeta_m^k )
		& = \sigma \left( \Im(\bq_m) \odot \psi\left( \tfrac{1}{\sigma} \Im(\bq_m) \odot \Im(\bz_m^k) \right)  \right),
	\end{align}
\end{subequations}
for all $m,n$, where $\odot$ denotes the element-wise product.
It can be verified that
the complexity of \eqref{eq:em1_r} and \eqref{eq:condmean_comp} is
\beq \label{eq:em_complex}
\bigO( 2MW (\log(W) + N/2 + C ) );
\eeq
particularly, we need FFTs and IFFTs for a total of $2M$ times.
The complexity in \eqref{eq:em_complex} is the per-iteration complexity of EM,  without counting the complexity required to solve \eqref{eq:em1_dec}.
We see that \eqref{eq:em_complex} is {\em not} much better than the per-iteration complexity \eqref{eq:pg_complex} for the PG method.
When the complexity of solving \eqref{eq:em1_dec} factors in, the per-iteration complexity of the EM method is likely to be higher than that of PG.
The next question then goes to whether EM leads to faster convergence.
We will try to answer this question
 in the subsequent sections.

The EM method for OMOD was studied in \cite{plabst2018efficient}, wherein the efficient subcarrier-decoupling implementation was proposed,
and GMAP was considered.
For GMAP, the problems in \eqref{eq:em1_dec} have closed-form solutions, and this advantage was exploited.


\section{Convergence Analysis and New Insights}
\label{sect:conv}

In the last section, we asked the question of whether the EM method would converge faster than the PG method in the application of OMOD.
In this section we perform convergence analysis with EM.
We will show that, in theory, the EM method should converge faster than the PG method.
However, this is not the most striking result.
The insight gained from our analysis will lead us to develop
an accelerated variant of the EM method, which converges even faster.
It will also enable us to consider inexact variants of the EM methods, which provides flexibility with implementations.
We will also expand our analysis to include the non-convex case; i.e., instances for which the penalty function $h$ is non-convex.

This section is organized as follows.
Section~\ref{sect:conv_prelim} gives the problem setup.
Sections~\ref{sect:conv_pg} and \ref{sect:conv_em} describe the convergence results for the PG and EM methods, respectively.
Section~\ref{sect:conv:insight} reveals the insight behind our EM convergence analysis.
Section~\ref{sect:conv_slow} describes a result that indicates that the PG and EM methods can converge slowly for high SNRs.
Taking insight from the EM convergence analysis, Section~\ref{sect:conv_aem} and \ref{sect:conv_iem} develop the accelerated and inexact EM methods, respectively.
Section~\ref{sect:niem} considers the convergence analysis for the non-convex case.
It is worth noting that, while we develop these results for OMOD, they can be applied to other QLR problems.


\subsection{Problem Setup and Preliminary Concepts}
\label{sect:conv_prelim}

Let us write down our main problem, problem \eqref{eq:main_prob}, again:
\beq \label{eq:ana_P}
\min_{\btheta \in \Rbb^n} F(\btheta):= f(\btheta) + h(\btheta),
\eeq
where
$f(\btheta) = \textstyle - \sum_{i=1}^m \log \Phi( \bb_i^\top \btheta )$
is convex differentiable;
$h: \Rbb^n \rightarrow \Rbb \cup \{ +\infty \}$ is proper, lower semi-continuous and convex.
We assume that problem \eqref{eq:ana_P} has an optimal solution,
and we will use the notation $\btheta^\star$ to denote an optimal solution to problem \eqref{eq:ana_P}.
 In our study, we will see problem \eqref{eq:ana_P} as a problem with a general $\bB$ and $h$.
Hence, we basically consider a class of QLR problems, not just OMOD.

We also recall some basic formulas.
The gradient $\nabla f(\btheta)$ can be written as
\beq  \label{eq:ana_df}
\nabla f(\btheta) = -\bB^\top \psi(\bB\btheta);
\eeq
where
$\bB = [~ \bb_1,\ldots,\bb_m ~]^\top$; $\psi(\bz) = (\psi(z_1),\ldots,\psi(z_m))$;
\beq
\label{eq:psi_def}
\psi(z)
=  \frac{e^{-z^2/2}}{\int_{-\infty}^z e^{-t^2/2} {\rm d}t }.
\eeq
The EM surrogate function of $f$ can be written as
\begin{align}
g(\btheta|\btheta') & = \tfrac{1}{2} \| \bz' - \bB\btheta \|^2 + {\rm const.}
\label{eq:ana_g}
\\
\bz' & = \bB \btheta' + \psi(\bB \btheta');
\label{eq:ana_z}
\end{align}
see \eqref{eq:Qtt}--\eqref{eq:exa_EMM3}.
We have $g(\btheta|\btheta') \geq f(\btheta)$ for all $\btheta,\btheta'$, and $g(\btheta'|\btheta') = f(\btheta')$ for all $\btheta'$.

Some concepts and notations should be introduced.
Let $\varphi: \Rbb^n \rightarrow \Rbb \cup \{ + \infty \}$, and let $\setX \subseteq \Rbb^n$.
The domain of $\varphi$ is defined as $\dom \varphi= \{ \btheta \in \Rbb^n | \varphi(\btheta) < +\infty \}$.
A differentiable $\varphi$ is said to have $L_\varphi$-Lipschitz continuous gradient on $\setX$ if
\[
\| \nabla \varphi(\btheta) - \nabla \varphi(\btheta') \| \leq L_\varphi \| \btheta - \btheta' \|,
~ \text{for all $\btheta, \btheta' \in \setX$}.
\]
Consider a minimization problem $\min_{\btheta \in \Rbb^n} \varphi(\btheta)$.
A point $\hat{\btheta} \in \Rbb^n$ is said to be a critical point of the problem if $\bzero \in \partial \varphi(\hat{\btheta})$, where $\partial \varphi({\btheta})$ is the limiting subdifferential of $\varphi$ at $\btheta$; see~\cite{li2020understanding} and the references therein.
A point $\hat{\btheta}$ is said to be an $\eps$-critical point of the problem if
${\rm dist}(\bzero, \partial \varphi(\hat{\btheta})) \leq \eps$,
where ${\rm dist}(\by,\setX) = \min_{\bx \in \setX} \| \by - \bx \|$ denotes the distance between $\by$ and $\setX$.
If $\varphi$ is convex, then any critical point of the problem is an optimal solution to the problem.
The notations $\sigma_{\rm max}(\bX)$ and $\sigma_{\rm min}(\bX)$ denote the largest and smallest singular values of $\bX$, respectively;
$\langle \cdot, \cdot \rangle$ denotes the inner product.


\subsection{Proximal Gradient}
\label{sect:conv_pg}

Consider the PG method for problem \eqref{eq:ana_P}:
\beq \label{eq:ana_PG}
\btheta^{k+1} = \prox_{\eta h}( \btheta^k - \eta \nabla f(\btheta^k)),
\quad k=0,1,2,\ldots,
\eeq
where we assume a constant step size $\eta > 0$.
The following result is well known.
\begin{Fact}[{see, e.g.,\cite{beck2017first}}]
\label{fact:PG_conv}
Consider problem \eqref{eq:ana_P} and the accompanied problem setup, and consider the PG method  \eqref{eq:ana_PG}.
Suppose that $f$ has $L_f$-Lipschitz continuous gradient on $\dom F$.
If the step size is chosen as $\eta = 1/L_f$, then
we have
\beq
F(\btheta^k)- F(\btheta^\star) \leq \frac{L_f}{k} \| \btheta^0 - \btheta^\star \|^2, \quad k \geq 1. \nonumber
\eeq
\end{Fact}
Our question is whether
$f$
possesses the desired Lipschitz continuous gradient property.
This appears to have not been answered before, and we show that the answer is yes.
\begin{Prop} \label{prop:Lf}
	The function $f$ in \eqref{eq:ana_P} has $L_f$-Lipschitz continuous gradient on $\Rbb^n$, where $L_f = \sigma_{\rm max}(\bB)^2$.	
	Consequently,	
	the PG method  \eqref{eq:ana_PG} with step size $\eta= 1/\sigma_{\rm max}(\bB)^{2}$ has the following convergence result for problem \eqref{eq:ana_P}:
	\beq \label{eq:ana_PG_con_alt}
	F(\btheta^k)- F(\btheta^\star) \leq \frac{\sigma_{\rm max}(\bB)^2}{k} \| \btheta^0 - \btheta^\star \|^2, \quad k \geq 1.
	\eeq
\end{Prop}

The proof of Proposition \ref{prop:Lf} is given in Appendix~\ref{app:prop1}.
Proposition \ref{prop:Lf} gives several implications.
First, it
confirms that the PG method guarantees convergence to the optimal solution with a convergence rate of at least $1/k$.
Second, it indicates how the convergence scales with the problem instance $\bB$; specifically, it scales with $\sigma_{\rm max}(\bB)^2$.
This will be useful in our comparison with the EM method later.
Third, it suggests how we can select the step size, specifically, by $\eta= 1/\sigma_{\rm max}(\bB)^{2}$.
In the OMOD application, we can show from \eqref{eq:ana_B} that
\beq \label{eq:sigB}
\sigma_{\rm max}(\bB)^2  = \frac{\sigma_{\rm max}(\cH)^2}{\sigma^2}
= \max_{w=1,\ldots,W} \frac{ \sigma_{\rm max}(\check{\bH}_w)^2 }{\sigma^2},
\eeq
which gives us a simple way to select the step size.
Note that the prior OMOD work \cite{Mirfarshbafan2020} selects the step size manually.

\subsection{Expectation Maximization}
\label{sect:conv_em}

Next, consider the EM method for problem \eqref{eq:ana_P}:
\beq \label{eq:ana_EM}
\btheta^{k+1} \in \arg \min_{\btheta \in \Rbb^n}  g(\btheta|\btheta^k) + h(\btheta),
~ k=0,1,2,\ldots
\eeq
We have the following convergence result.
\begin{Prop} \label{prop:EM}
	Consider problem \eqref{eq:ana_P} and the accompanied problem setup, and consider
	 the EM method \eqref{eq:ana_EM}.
	We have
	\beq \label{eq:ana_EM_con}
	F(\btheta^k)- F(\btheta^\star) \leq \frac{1}{k} \| \bB( \btheta^0 - \btheta^\star ) \|^2, \quad k \geq 1.
	\eeq
\end{Prop}
The proof of Proposition~\ref{prop:EM} is given in Appendix \ref{app:prop2}, and we will describe the insight behind Proposition~\ref{prop:EM} in the  next subsection.
Eq.~\eqref{eq:ana_EM_con} shows that
the EM method has a convergence rate of at least $1/k$---same as PG.
It is interesting to compare the convergence results of PG and EM.
Since
\beq \label{eq:gap}
\| \bB( \btheta^0 - \btheta^\star ) \|^2 \leq \sigma_{\rm max}(\bB)^2 \| \btheta^0 - \btheta^\star \|^2,
\eeq
the EM convergence result \eqref{eq:ana_EM_con} is seen to be at least no worse than the PG result in \eqref{eq:ana_PG_con_alt}.
In fact, \eqref{eq:gap} suggests that the EM method should converge faster than the PG method, particularly when the gap in \eqref{eq:gap} is large.


\subsection{Insight Behind the Convergence Results,  and Prior Work}
\label{sect:conv:insight}

It is interesting to see what is
the insight behind the EM convergence result in Proposition \ref{prop:EM}.
To see it, we need to understand
how PG convergence results were established in the literature.
The idea is to see 
PG
as
a majorization-minimization (MM)
method.
Specifically, consider the MM method
\beq \label{eq:ana_MM}
\btheta^{k+1} \in \arg \min_{\btheta \in \Rbb^n} u(\btheta|\btheta^k) + h(\btheta),
\quad k=0,1,2,\ldots
\eeq
where $u(\cdot|\btheta')$ is a majorant of $f$ at $\btheta'$; i.e., $u(\btheta|\btheta') \geq f(\btheta)$ for all $\btheta,\btheta'$,  $u(\btheta'|\btheta') = f(\btheta')$ for all $\btheta'$.
The PG method \eqref{eq:ana_PG} with step size $\eta = 1/L_f$ is seen as an MM, with
\beq \label{eq:ana_u_PG}
u(\btheta|\btheta') = f(\btheta') + \langle \nabla f(\btheta'), \btheta - \btheta' \rangle + \tfrac{L_f}{2} \| \btheta - \btheta' \|^2.
\eeq
Many PG convergence proofs
make strong use of the identity \eqref{eq:ana_u_PG}.
As an important discovery, we found that the EM surrogate function $g$ in \eqref{eq:ana_g}  possesses the following structure
\beq \label{eq:ana_u_EM}
g(\btheta|\btheta') = f(\btheta') + \langle \nabla f(\btheta'), \btheta - \btheta' \rangle + \tfrac{1}{2} \| \bB(\btheta - \btheta') \|^2.
\eeq
We see that \eqref{eq:ana_u_EM} resembles \eqref{eq:ana_u_PG}.
This has tremendous insights.
It suggests that we may use the PG convergence proof concepts in the literature to show the convergence for EM.
In fact, the proof of Proposition \ref{prop:EM} is conceptually the same as that of Fact~\ref{fact:PG_conv}.

In view of the importance of \eqref{eq:ana_u_EM}, we provide the derivations of \eqref{eq:ana_u_EM} here.
From \eqref{eq:ana_g}--\eqref{eq:ana_z}, we have
\begin{align*}
	& \| \bz' - \bB \btheta \|^2 = \| \psi(\bB\btheta') - \bB(\btheta-\btheta') \|^2 \\
	& ~ = \| \psi(\bB\btheta') \|^2 - 2 \langle \psi(\bB\btheta'), \bB(\btheta-\btheta') \rangle + \| \bB(\btheta-\btheta') \|^2  \\
	& ~ = \| \psi(\bB\btheta') \|^2 + 2 \langle \nabla f(\btheta'), \btheta-\btheta' \rangle + \| \bB(\btheta-\btheta') \|^2,
\end{align*}
where the last equation is due to $\nabla f(\btheta') = -\bB^\top \psi(\bB\btheta')$ in \eqref{eq:ana_df}. By setting $\btheta = \btheta'$ in the above equation, we get
\[
\| \bz' - \bB \btheta' \|^2 = \| \psi(\bB\btheta') \|^2.
\]
Applying the above two equations to \eqref{eq:ana_g}, and using $g(\btheta'|\btheta') = f(\btheta')$, we obtain  \eqref{eq:ana_u_EM}.
The proof is complete.

The prior works on EM convergence analyses should be mentioned and compared.
The question is whether the available convergence analysis results \cite{wu1983convergence,razaviyayn2013unified,mairal2013optimization,hong2017iteration} can lead to the same or better claim than our result in \eqref{eq:ana_EM_con}.
This aspect is complex, and we  provide a detailed account of it in the supplemental material. In summary, the best result we can find is
\beq \label{eq:con_MM2}
F(\btheta^k) - F(\btheta^\star) \leq \frac{\sigma_{\rm max}(\bB)^2}{k+1}
(2R^2),
\eeq
for some constant $R > 0$;
it comes from Mairal's analysis \cite{mairal2013optimization}.
Our result in \eqref{eq:ana_EM_con} appears to be better than \eqref{eq:con_MM2}; e.g., it is unconvincing to use \eqref{eq:con_MM2} to argue that EM should converge faster than PG.

\subsection{A Slow Convergence Phenomenon}
\label{sect:conv_slow}

The PG and EM theoretical bounds in \eqref{eq:ana_PG_con_alt} and
\eqref{eq:ana_EM_con} are seen to be loose if $\bB$ has a large magnitude.
From \eqref{eq:ana_B}, the magnitude of $\bB$ is seen to increase with the SNR; see also \eqref{eq:sigB}.
This gives an impression that, under high SNRs, the PG and EM methods could converge slowly.
Further analysis shows that this is true
 for our interested choices of $h$.
\begin{Prop} \label{prop:new_EM_PG}
	Consider problem \eqref{eq:ana_P} and the accompanied problem setup.
	Consider either the PG method  \eqref{eq:ana_PG} with $\eta=1/\sigma_{\rm max}(\bB)^2$, or the EM method \eqref{eq:ana_EM}.
	Suppose that $\btheta^k$ is $\eps$-optimal in the sense that
	\[
	\| \btheta^k - \btheta^\star \| \leq \eps
	\]
	for some optimal solution $\btheta^\star$ to problem \eqref{eq:ana_P} and for $\eps > 0$.
	\begin{enumerate}[(a)]
		\item Suppose that $h(\btheta)= \Ind_\setX(\btheta)$ for some non-empty convex closed set $\setX$. Then $k$ must satisfy
		\[
		k \geq \sigma_{\rm max}(\bB) \frac{ \|  \btheta^\star - \btheta^0 \| -\eps }{ \sqrt{m} }
		\]
		for the PG method; and
		\[
		k \geq \sigma_{\rm min}^+(\bB) \frac{ \| \bP( \btheta^\star - \btheta^0 ) \| -\eps }{ \sqrt{m} }
		\]
		for the EM method, where
		$\sigma_{\rm min}^+(\bB)$ denotes the smallest positive singular value of $\bB$, and 		
		 $\bP = \bB^\top (\bB^\top \bB)^\dag \bB$ is an orthogonal projection matrix, which performs projection onto the subspace spanned by $\bb_1,\ldots,\bb_m$.
		
		\item Suppose that $h(\btheta)=  \lambda \| \btheta \|^2/2$, $\lambda > 0$. Then $k$ must satisfy
		\beq\label{eq:GMAP_PG}
		k \geq \sigma_{\rm max}(\bB)^2 \frac{\|  \btheta^\star - \btheta^0 \| -\eps}{\lambda \| \btheta^0 \| + \sigma_{\rm max}(\bB) \sqrt{m} }
		\eeq
		for the PG method; and
		\beq\label{eq:GMAP_EM}
		k \geq (\sigma_{\rm min}^+(\bB))^2 \frac{ \| \bP( \btheta^\star - \btheta^0 ) \| -\eps }{ \lambda \| \bP \btheta^0 \| + \sigma_{\rm min}^+(\bB) \sqrt{m} }
		\eeq
		for the EM method.
		
	\end{enumerate}
\end{Prop}
 The proof of Proposition~\ref{prop:new_EM_PG} is given in Appendix \ref{app:prop3}.
Proposition~\ref{prop:new_EM_PG} shows that the number of iterations required by the PG or EM method must increase with the magnitude of $\bB$, or the SNR, assuming that the other variables are held fixed.
Let us show the insight behind Proposition~\ref{prop:new_EM_PG}.
As an example, consider $h(\btheta) = 0$ and the PG method  \eqref{eq:ana_PG} with step size $\eta= 1/\sigma_{\rm max}(\bB)^{2}$.
From \eqref{eq:ana_df} one can show that
\beq \label{eq:insight_new_EM_PG_1}
	\| \nabla f(\btheta) \| \leq \sigma_{\rm max}(\bB) \| \psi(\bB \btheta) \| \leq \sigma_{\rm max}(\bB) \sqrt{m},
\eeq
where the second inequality is due to $0 < \psi(z) < 1$ (see \eqref{eq:psi_def}).
Applying \eqref{eq:insight_new_EM_PG_1} to the PG method \eqref{eq:ana_PG} gives
\beq \label{eq:insight_new_EM_PG_2}
\| \btheta^{k+1} - \btheta^k \| = \eta \| \nabla f(\btheta^k) \| \leq \sigma_{\rm max}(\bB)^{-1} \sqrt{m}.
\eeq
We see that, if the magnitude of $\bB$ is large, then every $\btheta^k$ changes little, and we have slow convergence.
This is the idea that underlies the results in Proposition~\ref{prop:new_EM_PG}.

\subsection{Accelerated EM}
\label{sect:conv_aem}

The PG-EM
relationship
revealed in
 Section~\ref{sect:conv:insight}
opens new possibilities.
In first-order convex optimization, it is well known that
the convergence of the PG method can be accelerated by applying extrapolation
 \cite{nesterov2007gradient,beck2009fast}.
We can accelerate the EM method in the same manner.
Consider the following modified EM method
\begin{align}
\btheta^{k+1} & \in \arg \min_{\btheta \in \Rbb^n}  g(\btheta|{\bvartheta}^k) + h(\btheta),
\label{eq:ana_AEM}
\\
\bvartheta^{k+1} & = \btheta^{k+1} + \alpha_{k+1} (\btheta^{k+1}- \btheta^{k}),
\label{eq:ana_AEM_extra}
\end{align}
for $k \geq 0$,
where $\bvartheta^{0}= \btheta^0$; $\{ \alpha_k \}_{k \geq 1}$ is an extrapolation coefficient sequence.
We employ the
FISTA sequence
\beq \label{eq:ana_FISTA1}
\alpha_{k+1} = \frac{t_k -1}{t_{k+1}},
\eeq
where $t_0 = 1$,
\beq \label{eq:ana_FISTA2}
t_{k+1} = \frac{  1+ \sqrt{1+4t_{k}^2}}{2}, \quad k \geq 0.
\eeq
The modified EM method in \eqref{eq:ana_AEM}--\eqref{eq:ana_FISTA2} will be called the accelerated EM method, in what follows.
The accelerated EM method is the direct application of the extrapolation in accelerated PG to the standard EM method \eqref{eq:ana_EM}.
Note that,
if we replace $g(\btheta|\bvartheta^k)$ in \eqref{eq:ana_AEM} with $u(\btheta|\bvartheta^k)$ in \eqref{eq:ana_u_PG}, we return to the 
accelerated PG method.
 We have the following result.
\begin{Prop} \label{prop:AEM}
	Consider problem \eqref{eq:ana_P} and the accompanied problem setup, and consider
	the accelerated EM method \eqref{eq:ana_AEM}--\eqref{eq:ana_FISTA2}.
	We have
	\beq \label{eq:ana_AEM_con}
	F(\btheta^k)- F(\btheta^\star) \leq \frac{2}{(k+1)^2} \| \bB( \btheta^0 - \btheta^\star ) \|^2, \quad k \geq 1.
	\eeq 	
\end{Prop}
The proof of Proposition~\ref{prop:AEM} is given in Appendix \ref{app:prop4}.
Proposition~\ref{prop:AEM} indicates that accelerated EM has  an accelerated convergence rate of at least $1/k^2$, which is the same as that of the accelerated PG method.


\subsection{Inexact EM}
\label{sect:conv_iem}

The EM method \eqref{eq:ana_EM} requires us to
solve the problems in \eqref{eq:ana_EM} exactly.
When the exact solutions are computationally non-negligible to compute, we may want to accept economical, but inexact, solutions.
This motivates the study of inexact EM
\beq \label{eq:ana_IEM}
\btheta^{k+1} \approx \arg \min_{\btheta \in \Rbb^n}  G(\btheta|\btheta^k):= g(\btheta|\btheta^k) + h(\btheta),
\eeq
for $k \geq 0$,
where $\btheta^{k+1}$ is an approximate solution,
produced by a solver that may not give high solution precision due to its limited complexity.
The question is
how robust the EM method is under solution errors.

Intriguingly, the same kind of questions has been addressed in the context of inexact PG \cite{schmidt2011convergence}.
We take insight from the aforementioned context to answer the question at hand.
We characterize the inexactness of $\btheta^{k+1}$ by
\beq \label{eq:ana_IEM_eps}
{\rm dist}(\bzero, \partial G(\btheta^{k+1}|\btheta^k)) \leq \eps_{k+1},
\eeq
for some $\eps_{k+1} \geq 0$; $\eps_{k+1}= 0$ means that $\btheta^{k+1}$ is an optimal solution to the problem in \eqref{eq:ana_IEM}.
Also, we assume the following descent condition
\beq \label{eq:ana_IEM_desc}
G(\btheta^{k+1}|\btheta^k) \leq G(\btheta^{k}|\btheta^k).
\eeq
Eq.~\eqref{eq:ana_IEM_desc} makes sense in the following way: If
 we do a warm start with the inexact solver for problem  \eqref{eq:ana_IEM} by using $\btheta^k$ as the initialization,
the resulting approximate solution $\btheta^{k+1}$ should yield an objective value $G(\btheta^{k+1}|\btheta^k)$ better than the initial objective value $G(\btheta^{k}|\btheta^k)$.
We have the result below.

\begin{Prop}  \label{prop:IEM}
	Consider problem \eqref{eq:ana_P} and the accompanied problem setup. Consider
	the inexact EM method \eqref{eq:ana_IEM},
	with \eqref{eq:ana_IEM_eps}--\eqref{eq:ana_IEM_desc} being satisfied.
	Suppose that $\bB$ has full column rank.
	For $k \geq 1$, we have
	\beq \label{eq:ana_IEM_con} \nonumber
	F(\btheta^k)- F(\btheta^\star)
	\leq \frac{1}{2k} \left(  \| \bB( \btheta^0 - \btheta^\star ) \| + 2 \frac{\sum_{i=1}^k \eps_i }{\sigma_{\rm min}(\bB)} \right)^2.
	\eeq
\end{Prop}

The proof of Proposition~\ref{prop:IEM} is given in Appendix \ref{app:prop5}.
Proposition~\ref{prop:IEM} reveals two aspects.
First,
 the inexact EM method can lead to convergence, with a rate of at least $1/k$,
if the solution accuracy $\eps_k$ improves with $k$ in such a way that $\sum_{i=1}^\infty \eps_i < \infty$;
i.e., $\eps_k$ decreases at a rate of $k^{-p}$ for $p > 1$, or faster.
Second, the inexact EM method may be more sensitive to solution errors if the smallest singular value $\sigma_{\rm min}(\bB)$ of $\bB$ is smaller, i.e., $\bB$ is more ill-conditioned.
In massive MIMO, we should expect that $\bB$ is reasonably well-conditioned.

We can also consider the accelerated version of inexact EM.
\begin{Prop}  \label{prop:AIEM}
	Consider problem \eqref{eq:ana_P} and the accompanied problem setup. Consider the accelerated version of
	the inexact EM method \eqref{eq:ana_IEM},
	wherein $\btheta^k$ is replaced by $\bvartheta^k$ in \eqref{eq:ana_AEM_extra}--\eqref{eq:ana_FISTA2}.
	Suppose that \eqref{eq:ana_IEM_eps} holds, and that $\bB$ has full column rank.
	For $k \geq 1$, we have
	\beq \label{eq:ana_AIEM_con} \nonumber
	F(\btheta^k)- F(\btheta^\star)
	\leq \frac{2}{(k+1)^2} \left(  \| \bB( \btheta^0 - \btheta^\star ) \| + 2 \frac{\sum_{i=1}^k i \, \eps_i}{\sigma_{\rm min}(\bB)}  \right)^2.
	\eeq
\end{Prop}
The proof of Proposition~\ref{prop:AIEM} is given in Appendix \ref{app:prop6}.
Note that the above result does not require the descent condition \eqref{eq:ana_IEM_desc}.
Proposition \ref{prop:AIEM} looks similar to
its non-accelerated counterpart in Proposition \ref{prop:IEM}.
The notable difference is that, to achieve a convergence rate of at least $1/k^2$ in Proposition \ref{prop:AIEM}, we need the solution accuracies $\eps_k$'s to satisfy $\sum_{i=1}^\infty i \, \eps_i < \infty$; i.e.,
$\eps_k$ decreases at a rate of $k^{-p}$, $p > 2$, or faster.
This means that the accelerated inexact EM method may require higher solution accuracies than the non-accelerated counterpart.

\subsection{The Case of Non-Convex $h$}
\label{sect:niem}


Our preceding analyses assume convex $h$.
We also want to handle non-convex $h$.
Consider the same inexact EM scenario in \eqref{eq:ana_IEM}.
The new difficulty is that the problems in \eqref{eq:ana_IEM} are non-convex.
We assume that the inexact solver for the problems in  \eqref{eq:ana_IEM} can only guarantee finding an $\eps$-critical point, rather than an optimal or $\eps$-optimal solution.
The $\eps$-criticality of $\btheta^{k+1}$ is characterized by \eqref{eq:ana_IEM_eps}.
Also, we assume that the descent condition \eqref{eq:ana_IEM_desc} holds;
our justification is the same as before.
Furthermore, we modify
$G(\btheta|\btheta^k)$ in \eqref{eq:ana_IEM} as
\beq \label{eq:mod_G}
G(\btheta|\btheta^k)= g(\btheta|\btheta^k) + \tfrac{\tau}{2} \| \bB(\btheta - \btheta^k )\|^2  + h(\btheta)
\eeq
for some $\tau > 0$.
We have the following result.

\begin{Prop}  \label{prop:NIEM}
	Consider problem \eqref{eq:ana_P} and the accompanied problem setup,
	except that $h$ is non-convex.
	Consider
	the inexact modified EM method \eqref{eq:ana_IEM} and \eqref{eq:mod_G},
	with \eqref{eq:ana_IEM_eps}--\eqref{eq:ana_IEM_desc} being satisfied.
	Suppose that $\bB$ has full column rank.
	We have
	\beq \label{eq:ana_NEM_con}
	\min_{i=1,\ldots,k} {\rm dist}_{\bR}(\bzero,  \partial F(\btheta^{i})) \leq \frac{1}{\sqrt{k}}  C,
	\eeq
	where
	${\rm dist}_{\bR}(\by,\setX) = \min_{\bx \in \setX} \| \by - \bx \|_{\bR}$,
	$\| \bx \|_{\bR} = ( \bx^\top \bR \bx )^{1/2}$,
	$\bR = (\bB^\top \bB)^{-1}$, and
	\[
	C = 2
	\left( \frac{(2+\tau)^2}{\tau} ( F(\btheta^0)- F(\btheta^\star) ) +  \frac{\sum_{i=1}^k \eps_i^2}{2 \sigma_{\rm min}(\bB)^2}  \right)^{1/2}.
	\]	
\end{Prop}
The proof of Proposition~\ref{prop:NIEM} is given in Appendix \ref{app:prop7}.
Eq.~\eqref{eq:ana_NEM_con} employs an ellipsoidal extension of the $\eps$-critical point definition to pin down the convergence.
As an easier way to understand,
consider the following expression which can be shown to be an implication of \eqref{eq:ana_NEM_con}:
\beq \label{eq:ana_NEM_con_easy}
\min_{i=1,\ldots,k} {\rm dist}(\bzero, \partial F(\btheta^{i})) \leq \frac{1}{\sqrt{k}}  C \, \sigma_{\rm max}(\bB).
\eeq
Eq.~\eqref{eq:ana_NEM_con_easy} suggests that the EM method leads to a $\delta$-critical point in $\bigO(1/ \delta^2)$ iterations,
assuming that $\sum_{i=1}^\infty \eps_i^2 < \infty$.
\section{Accelerated EM Algorithms for OMOD}
\label{sect:AIEM}

In this section we use the new theoretical insights gained in the previous section to build efficient algorithms for OMOD.
Our attention is paid to the implementation of the accelerated inexact EM (AIEM) method.

\subsection{An AIEM Algorithm for a General Convex $h$}

Consider the OMOD problem \eqref{eq:main_prob} with a general convex $h$.
By modifying EM OMOD method in Section~\ref{sect:1d_mimo_ofdm} as the AIEM method in Section \ref{sect:conv_iem}, we have the following AIEM OMOD method:
Let $\bs^0$ be the initialization, and let $\bs_{\sf ex}^0 = \bs^0$.
We perform, for $k=0,1,2,\ldots$,
\begin{enumerate}[1.]
	\item E-step:
	For all $m$, compute the conditional mean estimate $\br_m^k = \Exp_{\br_m}[ \br_m | \bq, \bs_{\sf ex}^k]$ in the same way as \eqref{eq:condmean_comp}.

	\item M-step preparation:
	For all $m$, compute $\tilde{\br}_m^k = \bF \br_m^k$ via FFT.
	Let $\check{\br}_w^k = ( \tilde{r}_{1,w}^k,\ldots,\tilde{r}_{M,w}^k)$ for all $w$.

	\item M-step:
	For all $w$, compute the inexact solutions
	\beq \label{eq:aiem_m}
	\check{\bs}_w^{k+1} \approx \arg \min_{\bm{\mathsf{s}} \in \Cbb^N} \tfrac{1}{\sigma^2}
	\underbrace{\tfrac{1}{2} \| \check{\br}_w^k - \check{\bH}_w  \bm{\mathsf{s}} \|^2}_{:= \varphi_w^k(\bm{\mathsf{s}})} + h(\bm{\mathsf{s}}).
	\eeq
	Here
	 $\{ \check{\bs}_w^{k+1} \}_{w=1}^W$ satisfies a solution quality condition
	 \beq
	 \| \be^{k+1} \|   \leq \eps_{k+1},  \label{eq:aiem_sol_req}
	 \eeq
	 for some given $\eps_{k+1} > 0$, where $\be^{k+1} \in \Rbb^W$ is given by 	
	\begin{gather}
		e_w^{k+1}  = \min_{\bv \in \partial h(\check{\bs}_w^{k+1})} \| \sigma^{-2} \nabla \varphi_w^k(\check{\bs}_w^{k+1}) + \bv \|,
		~ \text{for all $w$.}
		\label{eq:aiem_sol_req2}
	\end{gather}

	\item Extrapolation for acceleration:
	\beq \label{eq:aiem_ex}
	\bs_{\sf ex}^{k+1} = \bs^{k+1} + \alpha_{k+1}( \bs^{k+1} - \bs^k ),
	\eeq
	where $\{ \alpha_k \}_{k \geq 1}$ is given by \eqref{eq:ana_FISTA1}--\eqref{eq:ana_FISTA2}.
\end{enumerate}
Note that if we solve the problems in \eqref{eq:aiem_m} exactly, the above method becomes the accelerated (exact) EM;
that if we set $\alpha_k = 0$ for all $k$, the above method becomes the regular non-accelerated EM;
and that \eqref{eq:aiem_sol_req} is identical to the solution quality requirement in \eqref{eq:ana_IEM_eps}.
Proposition~\ref{prop:AIEM} suggests that we can choose
$\eps_k = C \, k^{-p}$ for some $C > 0$ and $p > 2$.

We have not specified the inexact solver for the M-step problems in \eqref{eq:aiem_m}.
The AIEM method provides us with the freedom to choose the solver, and
in this study we employ the accelerated PG (APG) method.
Fixing $k$ and $w$, the APG method for the problem in \eqref{eq:aiem_m} is given by
\begin{align*}
	\bm{\mathsf{s}}^{j+1}_w & = \prox_{\eta_j \sigma^2 h}( \bm{\mathsf{s}}^j_{{\sf ex},w} - \eta_w \nabla \varphi_w^k(\bm{\mathsf{s}}^j_{{\sf ex},w})), \\
	\bm{\mathsf{s}}^{j+1}_{{\sf ex},w} & = \bm{\mathsf{s}}^{j+1}_w + \alpha_{j+1} ( \bm{\mathsf{s}}^{j+1}_w - \bm{\mathsf{s}}^{j}_w ),
\end{align*}
for $j \geq 0$,
where $\eta_w = 1/\sigma_{\rm max}(\check{\bH}_w)^2$ is the step size, and
$\nabla \varphi_w^k(\bm{\mathsf{s}}) = \check{\bH}_w^\her \check{\bH}_w \bm{\mathsf{s}} - \check{\bH}_w^\her \check{\br}_w^k$.
We stop the APG loop when $\{ \bm{\mathsf{s}}^{j+1}_w \}_{w=1}^W$ meets the solution quality requirement \eqref{eq:aiem_sol_req}.
Assembling the above components together, we obtain the AIEM algorithm in Algorithm~\ref{alg:aiem}.

\subsection{Algorithms for Specific $h$ and Some Implementation Details}

The AIEM algorithm in the last subsection is developed for a general convex $h$.
Let us consider the box case $h(\btheta) = \Ind_{[-U,U]^n}(\btheta)$.
The $\prox$ operation is given by the clipping function \eqref{eq:clip}.
Eq.~\eqref{eq:aiem_sol_req2}, which measures the solution quality (see also Line 16 of Algorithm~\ref{alg:aiem}), can be shown to be
\beq \label{eq:check_err}
e_w^{k+1} =
\| \chi( \tilde{\bg}_w^{k+1}, \breve{\bs}_w^{k+1} ) \|,
\eeq
where
$\chi(\bg,\btheta) = ( \chi(g_1,\theta_1),\ldots,\chi(g_n,\theta_n) )$;
$\chi(g,\theta) = |g|$ if $g\theta \geq 0$ or $|\theta| < U$, and $\chi(g,\theta) = 0$ otherwise;
$\tilde{\bg}_w^{k+1}=  ( \Re(  \sigma^{-2} \nabla \varphi_w^k(\check{\bs}_w^{k+1}) ), \Im( \sigma^{-2} \nabla \varphi_w^k(\check{\bs}_w^{k+1}) ) )$;
$\breve{\bs}_w^{k+1} = ( \Re( \check{\bs}_w^{k+1} ), \Im( \check{\bs}_w^{k+1} ) )$.
We relegate the proof of \eqref{eq:check_err} to the supplemental material.

Let us also consider the GMAP case $h(\btheta) = \lambda \| \btheta \|^2/2$.
The problems in \eqref{eq:aiem_m} can actually be solved in closed form \cite{plabst2018efficient}:
\beq \label{eq:GMAP_sol}
\check{\bs}_w^{k+1} = ( \check{\bH}^\her_w \check{\bH}_w + \lambda \sigma^2 \bI )^{-1} \check{\bH}^\her_w \check{\br}_w^k.
\eeq
Hence we can replace the APG solver (Line 10--17 of Algorithm~\ref{alg:aiem}) with \eqref{eq:GMAP_sol}, and
the result is an accelerated EM.


\begin{algorithm}[!t]
	\caption{AIEM for problem \eqref{eq:main_prob} with convex $h$} \label{alg:aiem}
	\begin{algorithmic}[1]
		
		\STATE \textbf{input:} $\bs^0$, $\{ \alpha_k \}_{k \geq 1}$ in \eqref{eq:ana_FISTA1}--\eqref{eq:ana_FISTA2}, $\{ \eps_k \}_{k \geq 1}$
		
		\STATE set $\bs_{\sf ex}^0 = \bs^0$, $\eta_w = 1/\sigma_{\rm max}(\check{\bH}_w)^2$ for all $w$ 
		
		\FOR {$k=0,1,2,\ldots$}
		
		\STATE compute $\bz_m^k = \bF^\her (\sum_{n=1}^N \bD_{m,n} \bs_{{\sf ex},n}^k )$ for all $m$
		
		\STATE compute $\bzeta_m^k$ for all $m$ via
		\begin{align*}
			\Re(\bzeta_m^k) & =  \sigma \left( \Re(\bq_m) \odot \psi\left( \tfrac{1}{\sigma} \Re(\bq_m) \odot \Re(\bz_m^k) \right)  \right) \\
			\Im(\bzeta_m^k) & =  \sigma \left( \Im(\bq_m) \odot \psi\left( \tfrac{1}{\sigma} \Im(\bq_m) \odot \Im(\bz_m^k) \right)  \right)
		\end{align*}
		
		\STATE set $\br_m^k = \bz_m^k + \bzeta_m^k$ for all $m$
		
		\STATE compute $\tilde{\br}_m^k = \bF \br_m^k$ for all $m$
		\STATE set $\check{\br}_w^k = ( \tilde{r}_{1,w}^k,\ldots,\tilde{r}_{M,w}^k)$ for all $w$
		\STATE set $\bm{\mathsf{s}}^0_w = \check{\bs}_w^k$, $\bm{\mathsf{s}}^0_{{\sf ex},w} = \bm{\mathsf{s}}^0_w$ for all $w$, set $j= 0$
		
		\REPEAT
		
		
		\STATE $\bm{\mathsf{g}}_w^j = \check{\bH}_w^\her \check{\bH}_w \bm{\mathsf{s}}^j_{{\sf ex},w} - \check{\bH}_w^\her \check{\br}_w^k$ for all $w$
		
		\STATE $\bm{\mathsf{s}}^{j+1}_w = \prox_{\eta_w \sigma^2 h}( \bm{\mathsf{s}}^j_{{\sf ex},w} - \eta_w \bm{\mathsf{g}}_w^j )$  for all $w$
		
		\STATE $\bm{\mathsf{s}}^{j+1}_{{\sf ex},w} = \bm{\mathsf{s}}^{j+1}_w + \alpha_{j+1} ( \bm{\mathsf{s}}^{j+1}_w - \bm{\mathsf{s}}^{j}_w )$ for all $w$
		
		\STATE $j=j+1$
		
		\STATE $\tilde{\bm{\mathsf{g}}}_w^j = \sigma^{-2}( \check{\bH}_w^\her \check{\bH}_w \bm{\mathsf{s}}^j_{w} - \check{\bH}_w^\her \check{\br}_w^k)$ for all $w$
		
		\STATE $e_w^j = \min_{\bv \in \partial h(\bm{\mathsf{s}}^{j}_{w})  } \| \tilde{\bm{\mathsf{g}}}_w^j + \bv \|$ for all $w$
		
		
		
		\UNTIL{$\sum_{w=1}^W (e_w^j)^2 \leq \eps_{k+1}^2$}
		
		\STATE set $\bs^{k+1}$ such that $\check{\bs}^{k+1}_w = \bm{\mathsf{s}}^{j}_w$ for all $w$
		
		\STATE $\bs_{\sf ex}^{k+1} = \bs^{k+1} + \alpha_{k+1}( \bs^{k+1} - \bs^k )$
		\STATE if a stopping rule is met, exit and {\bf output} $\bs^{k+1}$
		
		\ENDFOR
		
	\end{algorithmic}
\end{algorithm}

\section{A Deep Inexact EM Algorithm for OMOD}
\label{sect:DIEM}

In this section we apply deep unfolding \cite{gregor2010learning,Samuel2019learning} to the inexact EM algorithm to build a data-driven detector for OMOD.  
The idea of deep unfolding is to alter an existing iterative algorithm by  data-driven learning, thereby hoping to learn a better algorithm.
Specifically, we see each iteration of the predecessor algorithm as a network layer, untie some of the parameters or alter some of the structures, and  learn the untied parameters or new structures from data.
We tried various ways to deep-unfold Algorithm~\ref{alg:aiem}.
The successful one, by our empirical tests, is shown in Algorithm~\ref{alg:diem}; we will call the algorithm {\it DIEM}.
The untied parameters are marked red, and some notable changes marked blue.

\begin{algorithm}[!t]
	\caption{DIEM, inspired by AIEM in Algorithm~\ref{alg:aiem}} \label{alg:diem}
	\begin{algorithmic}[1]
		
		\STATE set $\bs^0 = \bzero$, $\bs_{\sf ex}^0 = \bs^0$

		\FOR {$k=0,1,2,\ldots,K-1$}
		
		\STATE compute $\bz_m^k = \bF^\her (\sum_{n=1}^N \bD_{m,n} {\blue \bs_{n}^k} )$ for all $m$
		
		\STATE compute $\bzeta_m^k$ for all $m$ via
		\begin{align*}
		\Re(\bzeta_m^k) & =  (\sigma + {\red \beta_k}) \left( \Re(\bq_m) \odot \psi\left( \tfrac{1}{\sigma} \Re(\bq_m) \odot \Re(\bz_m^k) \right)  \right) \\
		\Im(\bzeta_m^k) & =  (\sigma + {\red \beta_k}) \left( \Im(\bq_m) \odot \psi\left( \tfrac{1}{\sigma} \Im(\bq_m) \odot \Im(\bz_m^k) \right)  \right)
		\end{align*}
		
		\STATE set $\br_m^k = \bz_m^k + \bzeta_m^k$ for all $m$
		
		\STATE compute $\tilde{\br}_m^k = \bF \br_m^k$ for all $m$
		\STATE set $\check{\br}_w^k = ( \tilde{r}_{1,w}^k,\ldots,\tilde{r}_{M,w}^k)$ for all $w$
		
		\STATE $\bm{\mathsf{g}}_w^k = \check{\bH}_w^\her \check{\bH}_w \bm{\mathsf{s}}^{k}_{{\sf ex},w}
		- \check{\bH}_w^\her \check{\br}_w^k$ for all $w$
		
		\STATE
		$\bm{\mathsf{s}}^{k+1}_w =
		{\blue \Omega}_{{\red \gamma_k}}(
		\bm{\mathsf{s}}^{k}_{{\sf ex},w}
		- {\red \eta_k} \bm{\mathsf{g}}_w^j )$  for all $w$
		
		\STATE 
		$\bm{\mathsf{s}}^{k+1}_{{\sf ex},w} = \bm{\mathsf{s}}^{k+1}_{w} +  {\red \alpha_{k+1}} ( \bm{\mathsf{s}}^{k+1}_w - \bm{\mathsf{s}}^{k}_w )$ for all $w$
		
		\STATE set $\bs^{k+1}$ such that $\check{\bs}^{k+1}_w = \bm{\mathsf{s}}^{k+1}_w$ for all $w$

		\ENDFOR
		
	\end{algorithmic}
\end{algorithm}

The most notable alternation of DIEM is Line 9 of Algorithm~\ref{alg:diem}.
We replace the proximal operator $\prox_{\eta_w \sigma^2 h}$ (Line 12 of AIEM) with
a nonlinear activation function $\Omega_{\gamma}$.
We use
a multilevel sigmoid function;
its real-valued scalar version is
\beq \label{eq:Omega}
\Omega_{\gamma}(\theta) = \sum_{\mu \in \{ 0,\pm 2, \ldots, \pm 2(D-1) \}} \varrho(\gamma(\theta-\mu)),
\eeq
where $\gamma > 0$, $\varrho: \Rbb \rightarrow [-1,1]$ is a $0$-centered sigmoid function:
\beq \nonumber
\varrho(\theta) = \frac{2}{1 + e^{-\theta}} - 1;
\eeq
we have $ \Omega_{\gamma}(\bm{\mathsf{s}})  = [ \Re( \Omega_{\gamma}(s_i) + \jj \, \Im( \Omega_{\gamma}(s_i) ) ]_i$ for the complex-valued vector case.
The sigmoid functions were used in deep MIMO detection \cite{corlay2018multilevel} as the activation function.
The intuitive idea is that $\Omega_{\gamma}$ approaches the decision function of the constellation set as $\gamma \rightarrow \infty$,
and we want to use it to encourage the symbol estimates to lie closer to the constellation points.

The other alterations of DIEM are as follows:
we abandon acceleration in the EM loop;
we consider one-step APG with the inexact solver;
we untie the step size $\eta_k$ and extrapolation coefficient $\alpha_k$ of the APG in a layer-varying fashion;
we add a parameter $\beta_k$ to partially alter the noise standard deviation.

The training of DIEM is standard.
Let $\bp= \{ \bs, \cH, \sigma, \bq \}$ be a problem instance,
let $\bpi = \{ \balp, \bbeta, \bm\gamma, \bm\eta \}$ be the set of network parameters, and let $\bm{o}_\bpi(\bp)$
be the network output.
We generate a large number of problem instances $\bp_1,\ldots,\bp_T$ according to the signal model.
Then we learn $\bpi$ from $\bp_1,\ldots,\bp_T$ using a deep learning software; the objective is
\[ \textstyle
\min_{\bpi} \sum_{t=1}^T \| \bs_t - \bm{o}_\bpi(\bp_t) \|^2,
\]
where $\bs_t$ is the symbol vector of the problem instance $\bp_t$.

We want to provide an interpretation for DIEM.
Suppose
we choose a non-convex penalty 
$h$ with the OMOD \eqref{eq:main_prob} to better approximate the difficult constellation constraints.
As a concurrent study, we recently showed that there exists a constellation-promoting penalty function $h$ such that its proximal operator may be approximated by the sigmoid activation function \eqref{eq:Omega};
see \cite{shao2023explanation} for details.
As studied in Section~\ref{sect:niem}, the corresponding non-convex OMOD problem \eqref{eq:main_prob} can be handled by inexact EM.
Following Section~\ref{sect:AIEM},
we can show that the non-convex inexact EM can be implemented by
nearly the same procedure as the non-accelerated version of Algorithm~\ref{alg:aiem}.
From this view, DIEM can be regarded as the deep unfolding of the non-convex inexact EM.


\section{Numerical Results}
\label{sec:sim}

In this section numerical results are provided to illustrate the detection and runtime performance of the new algorithms.
We also give additional numerical results in the supplemental material.

\subsection{Simulation Settings}

We use Monte Carlo simulations to test the performance of the algorithms.
The signals are generated by the model in Section~\ref{sect:omod_model}.
The channel is generated by a multipath model; see, e.g., \cite{heath2016overview} and the reference therein.
To describe the model,
let $\bar{\bh}_{l,n}= (h_{l,1,n},\ldots,h_{l,M,n})$, $l=0,\ldots,W'-1$, be the $l$th coefficient of the time-domain channel impulse response from user $k$ to all the BS antennas.
Assuming that the BS antennas are arranged as a uniform linear array,
we model each $\bar{\bh}_{l,n}$ as
\[
\bar{\bh}_{l,n} = \textstyle \sum_{j=1}^J \alpha_{j,l,n} \ba(\theta_{j,l,n}),
\]
where $\ba(\theta) = ( 1, e^{-\jj \frac{2\pi d}{\lambda} \sin(\theta)},\ldots, e^{-\jj (M-1)\frac{2\pi d}{\lambda} \sin(\theta)})$ is the response of the array to a signal coming from an angle  $\theta \in (-\pi/2,\pi/2)$; $d$ is the inter-antenna spacing; $\lambda$ is the wavelength;
$J$ is the number of paths;
$\alpha_{j,l,n}$ and $\theta_{j,l,n}$ are the complex channel gain and angle of a path, respectively.
In our simulations, we set $W'= 16$, $J = 4$ and $d= \lambda/2$;
each $\alpha_{j,l,n}$ is randomly generated, following a circular complex Gaussian distribution with mean zero and variance $1/J$;
each $\theta_{j,l,n}$ is randomly generated, following a uniform distribution on $(-\pi/2,\pi/2)$.
The number of trials of our simulations is $500$.
In addition, the SNR is defined as ${\sf SNR} = \Exp[ \| \bs \|^2 ]/\Exp [ \| \bv_n \|^2 ]$.

Our simulations consider the following algorithms.
\begin{enumerate}[1.]
	\item GMAP-EM: the standard EM method for GMAP OMOD, proposed previously in \cite{plabst2018efficient} and reviewed in Section~\ref{sect:1d_mimo_ofdm};
	\item GMAP-AEM: the accelerated EM method for GMAP OMOD, built in Section~\ref{sect:AIEM};
	\item BOX-PG: the PG method for box OMOD, proposed previously in \cite{Mirfarshbafan2020} and reviewed in Section~\ref{sect:1c_mimo_ofdm};
	\item BOX-AIEM: the accelerated inexact EM method for box OMOD, built in Section~\ref{sect:AIEM};
	\item BOX-EM: the standard EM method for box OMOD, implemented by BOX-AIEM with a high solution accuracy;
	\item DIEM: the deep algorithm in Section~\ref{sect:DIEM};
	\item ZF: direct application of the zero-forcing detector in unquantized MIMO-OFDM to OMOD.
\end{enumerate}
We tested all the algorithms by MATLAB 8.5 on the same desktop with Intel i7-7700 processor and 16GB RAM memory.

The parameter settings with the PG and EM algorithms (the non-deep ones) are as follows.
The stopping condition is either
$\| \bs^{k+1} - \bs^k \|/\| \bs^k \| \leq 5 \times 10^{-4}$ or $k > 1000$.
For BOX-AIEM, we set the M-step solution accuracies as $\eps_k = 2 N W k^{-2.1}$.
For BOX-EM, we set $\eps_k = 2 N W \times 10^{-4}$.
For BOX-PG, we choose the step size $\eta$ as the reciprocal of \eqref{eq:sigB}; see Proposition~\ref{prop:Lf}.
We set the noise power parameter $\sigma^2$ as
\beq \label{eq:loading}
\sigma = \sigma_{\rm actual} + \sigma_0,
\eeq
where $\sigma_{\rm actual}^2$ is the actual noise power;
$\sigma_0 > 0$ is some given constant.
Doing so is to avoid slow convergence.
As shown in Proposition~\ref{prop:new_EM_PG}, the PG and EM methods can converge slowly for small $\sigma^2$.
In our simulations, we set $\sigma_0 = 3$.

The training
 of DIEM
is implemented on Pytorch 1.2 platform for Python 3.5.2  (the test of DIEM is on MATLAB).
The training optimizer is stochastic gradient descent, with  step size $10^{-3}$.
The SNR range in our training is the SNR range in our simulations, with an additional margin of $2$ to $5$dB.
The number of layers is $K= 20$.
 The open source code is available at \url{https://github.com/shaomingjie/DeepEM_MIMO_det.git}.


\begin{figure}
  \centering
  \includegraphics[width=0.7\linewidth]{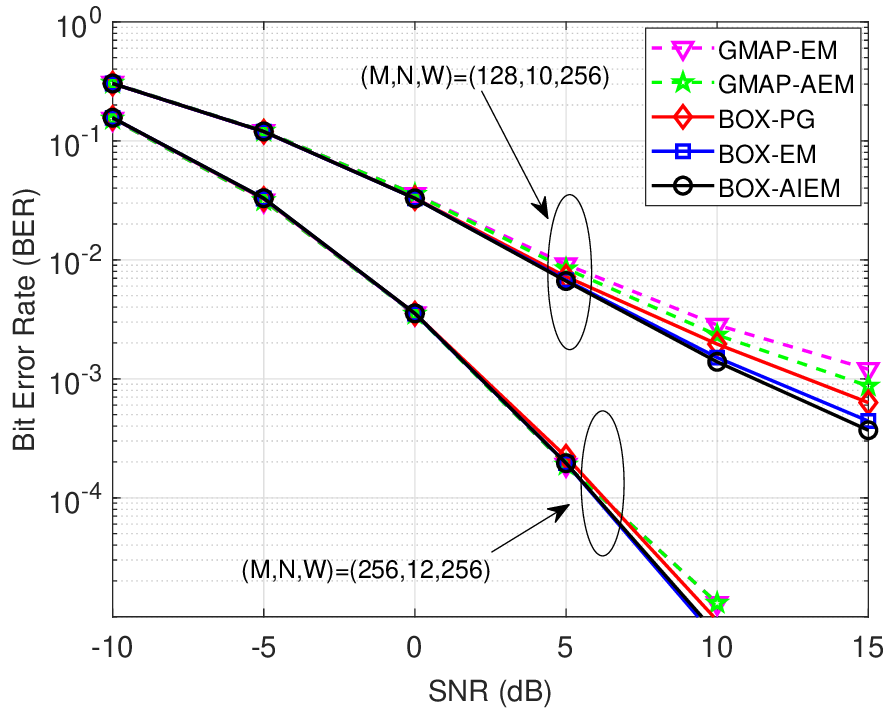}
  \caption{BERs of the PG and EM algorithms. $16$-QAM.  Dashed: GMAP-based algorithms; solid: box-based algorithms.}\label{fig:BER_GMAP_BOX}
\end{figure}

\begin{figure}[htb!]
	\centering
	\begin{subfigure}{0.48\textwidth}
		\centering
		\includegraphics[width=\linewidth]{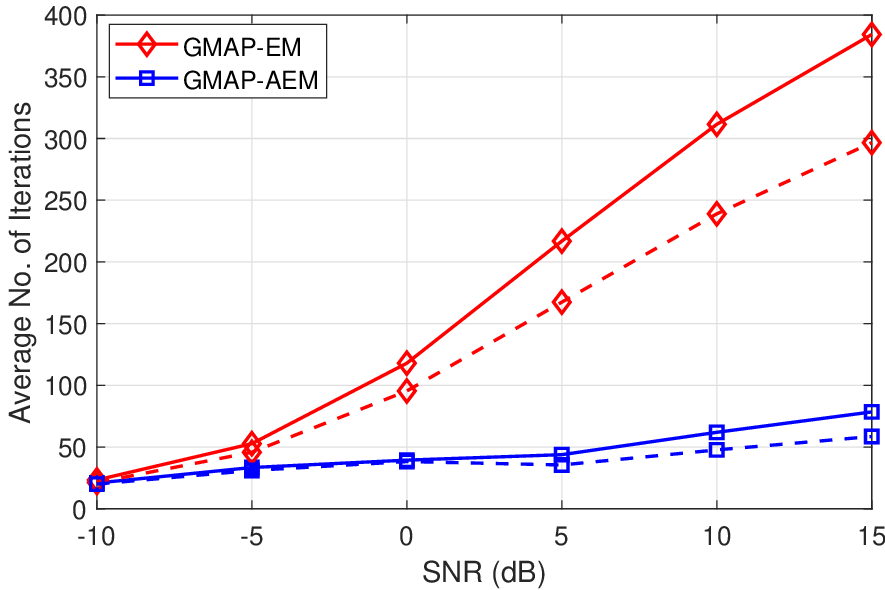}
		\caption{GMAP}
	\end{subfigure}
	\quad
	\begin{subfigure}{0.48\textwidth}
		\centering
		\includegraphics[width=\linewidth]{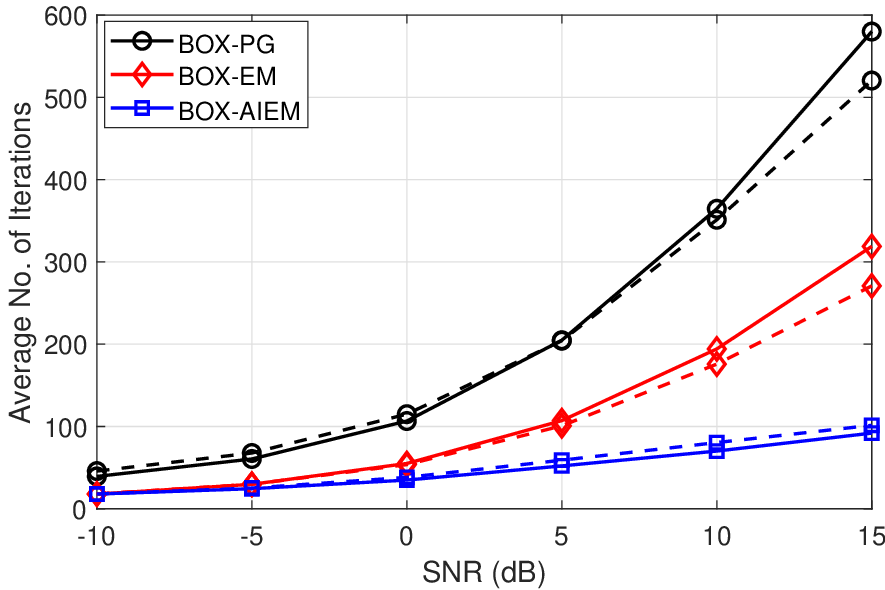}
		\caption{Box}
	\end{subfigure}
	\caption{Average number of iterations of the PG and EM algorithms. 16-QAM.  Solid lines: $(M,N,W) = (256,12,256)$, dashed lines: $(M,N,W) = (128,10,256)$.}\label{fig: iter_GMAP_BOX}
\end{figure}

\begin{figure}[htb!]
	\centering
	\begin{subfigure}{0.48\textwidth}
		\centering
		\includegraphics[width=\linewidth]{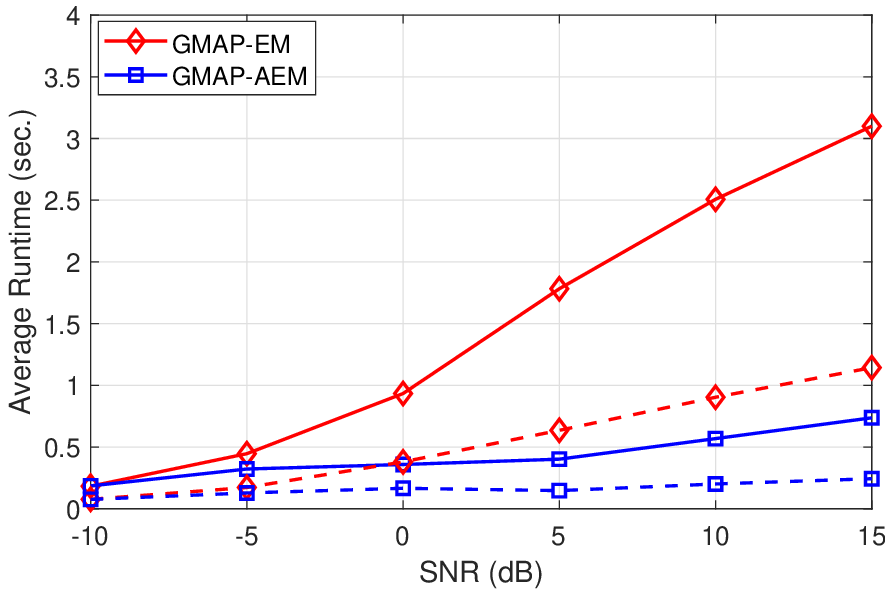}
		\caption{GMAP}
	\end{subfigure}
	\quad
	\begin{subfigure}{0.48\textwidth}
		\centering
		\includegraphics[width=\linewidth]{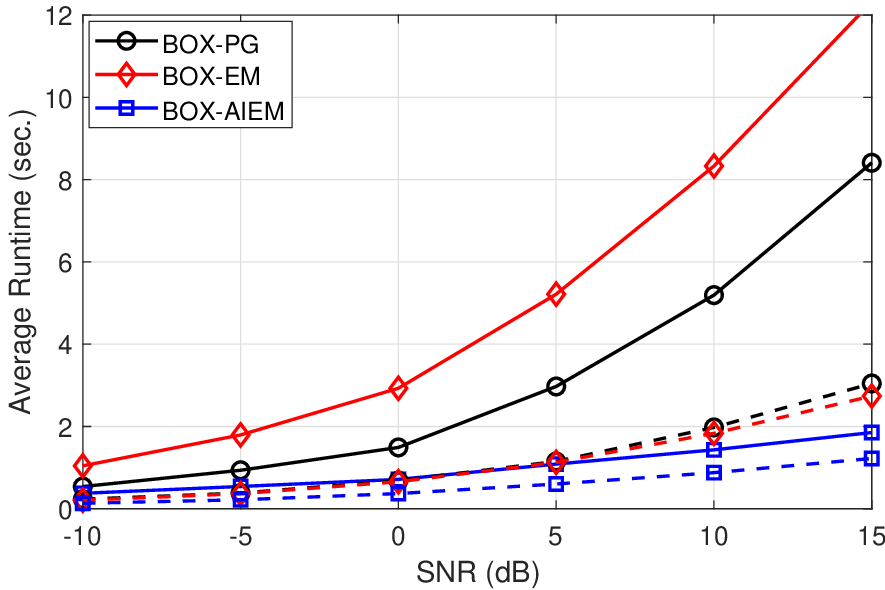}
		\caption{Box}
	\end{subfigure}
	\caption{Average runtimes of the PG and EM algorithms.  16-QAM.  Solid lines: $(M,N,W) = (256,12,256)$, dashed lines: $(M,N,W) = (128,10,256)$. }\label{fig: runtime_GMAP_BOX}
\end{figure}

\subsection{Comparison of EM and Accelerated Exact/Inexact EM}

In this subsection we focus on comparison of the standard EM method and our modified EM methods.
Fig.~\ref{fig:BER_GMAP_BOX} shows the bit error rate (BER) performance of the GMAP- and box-based algorithms.
GMAP-AEM is seen to yield nearly the same performance as GMAP-EM.
This is expected since both solve the same problem, and the problem is convex.
Similarly, we see the same behaviors with the box-based algorithms.
Fig.~\ref{fig: iter_GMAP_BOX} shows the average numbers of EM or PG iterations used by the algorithms.
As can be seen, the accelerated EM methods use much less iterations than the PG and standard EM methods.
This demonstrates the benefit of accelerated EM.

From Fig.~\ref{fig: iter_GMAP_BOX}, we also observe a phenomenon, namely, that the numbers of iterations of the PG and EM algorithms rise with the SNR.
This agrees with the theoretical results in Proposition~\ref{prop:new_EM_PG}.

Fig.~\ref{fig: runtime_GMAP_BOX} shows the average runtimes.
It is more interesting to examine the box case.
While we see in Fig.~\ref{fig: iter_GMAP_BOX}(b) that BOX-EM uses less numbers of iterations than BOX-PG,
we see in Fig.~\ref{fig: runtime_GMAP_BOX}(b) that BOX-EM runs slower than BOX-PG.
This is because the M-step in EM requires non-negligible computations.
Moreover, BOX-AIEM is seen to be faster than BOX-EM (and also BOX-PG); the use of inexact M-step is a key factor for its runtime advantage.

\subsection{Performance Comparison for DIEM}

\begin{figure}[htb!]
	\centering
	\begin{subfigure}{0.48\textwidth}
		\centering
		\includegraphics[width=\linewidth]{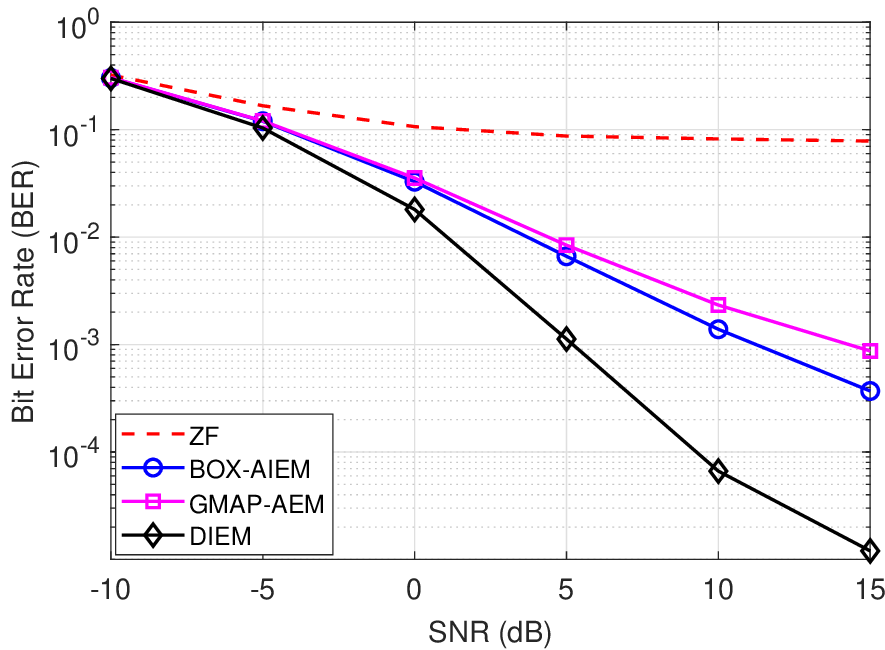}
		\caption{$(M,N,W)=(128,10,256)$}
	\end{subfigure}
	\quad
	\begin{subfigure}{0.48\textwidth}
		\centering
		\includegraphics[width=\linewidth]{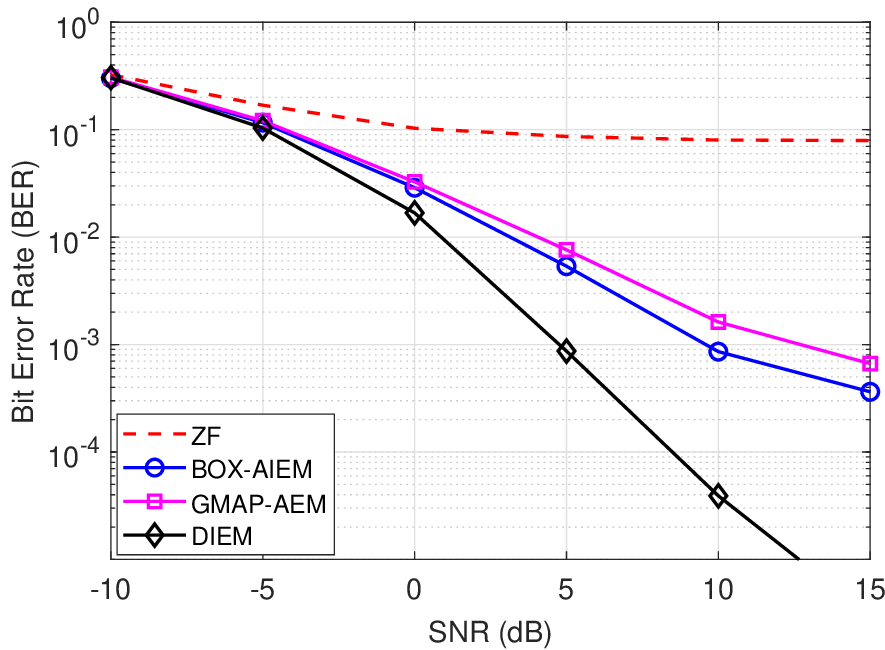}
		\caption{$(M,N,W)=(256,20,512)$}
	\end{subfigure}
	\caption{ BER performance of DIEM; 16-QAM.  }\label{fig: BER_16QAM}
\end{figure}

\begin{figure}[htb!]
	\centering
	\begin{subfigure}{0.48\textwidth}
		\centering
		\includegraphics[width=\linewidth]{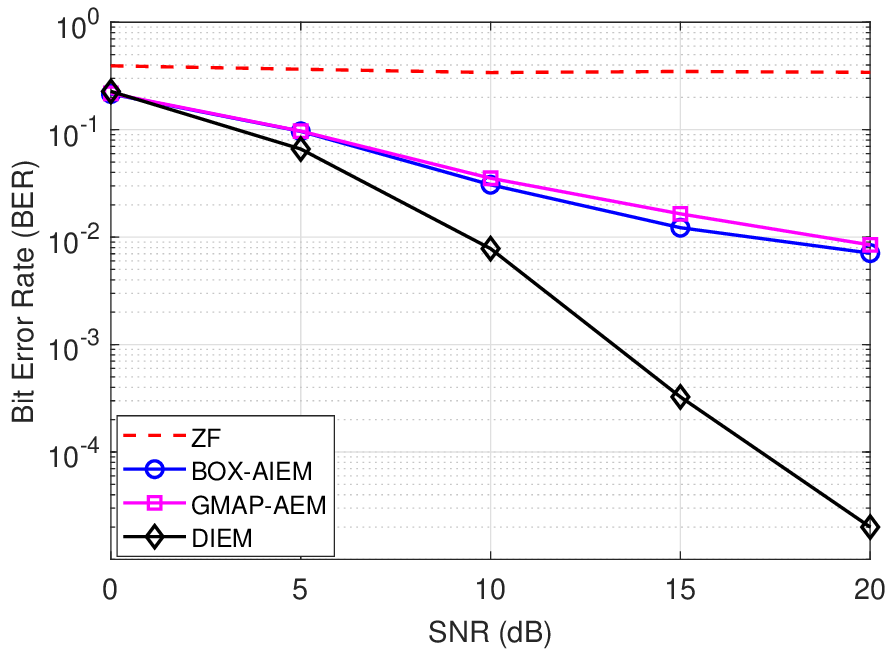}
		\caption{$(M,N,W)=(128,6,256)$}
	\end{subfigure}
	\quad
	\begin{subfigure}{0.48\textwidth}
		\centering
		\includegraphics[width=\linewidth]{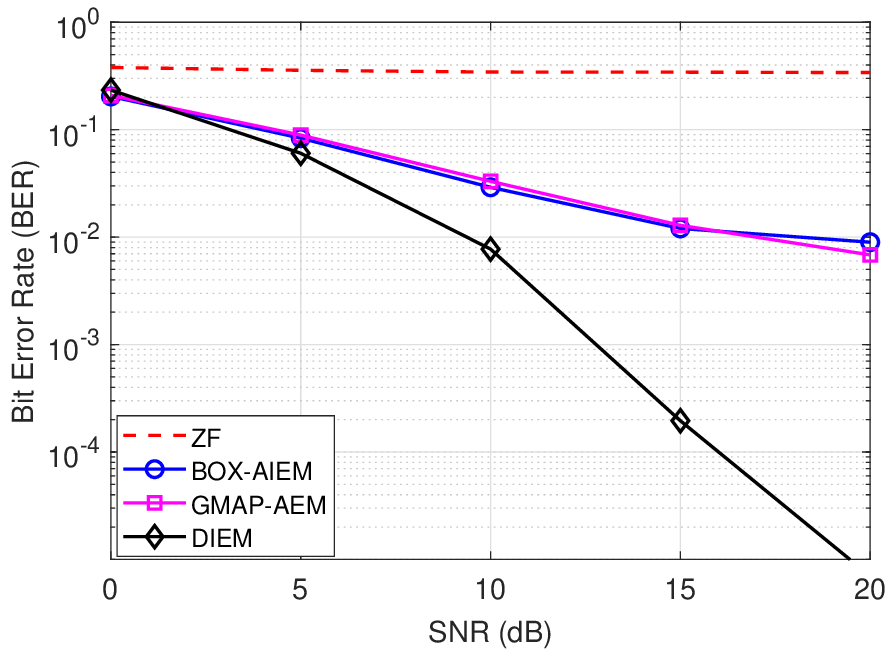}
		\caption{$(M,N,W)=(256,12,512)$}
	\end{subfigure}
	\caption{ BER performance of DIEM; 64-QAM. }\label{fig: BER_64QAM}
\end{figure}

In this subsection we demonstrate the performance of DIEM.
Figs.~\ref{fig: BER_16QAM}--\ref{fig: BER_64QAM} show the BER performance of DIEM under various system settings.
We benchmark DIEM against ZF, GMAP-AEM and BOX-AIEM.
DIEM is seen to exhibit remarkably better BER performance than the other algorithms.
Tables~\ref{tb:DIEM_16QAM}--\ref{tb:DIEM_64QAM} show the runtimes of the various algorithms.
As can be seen, DIEM is much faster than the other algorithms.
The reason for the efficiency of DIEM is that it uses a fixed number of layers, specifically, $20$; and each layer has simple operations, specifically, one-step APG-like operations.

\begin{table}[htb!]
	\centering
	\caption{Average Runtimes of   different algorithms. $(M,N,W)=(128,10,256)$, $16$-QAM }\label{tb:DIEM_16QAM}
	\renewcommand{\arraystretch}{1.2}
	\resizebox{0.7\linewidth}{!}{%
		\begin{tabular}{M{37mm}|M{8mm}|M{8mm}| M{8mm} |M{8mm}|M{8mm}|M{8mm}}
			\hline  
			\backslashbox{Alg.}{SNR (dB)}   &   -10      &   -5    & 0 & 5 & 10 & 15 \\
			\hline \hline 
			GMAP-AEM & 0.076 &	0.13 &	0.17 &	0.23&	0.28&	0.33  \\
			\hline
			
			BOX-AIEM & 0.13 &	0.20 &	0.37 &	0.60 & 0.87 & 1.22 \\
			
			\hline 
			
			DIEM &  \multicolumn{6}{c}{0.094}\\
			\hline
		\end{tabular}
	}
\end{table}

\begin{table}[htb!]
	\centering
	\caption{Average Runtimes of   different algorithms. $(M,N,W)=(128,6,256)$, $64$-QAM }\label{tb:DIEM_64QAM}
	\renewcommand{\arraystretch}{1.2}
	\resizebox{0.7\linewidth}{!}{%
		\begin{tabular}{M{37mm}|M{10mm}|M{10mm}| M{10mm} |M{10mm}|M{10mm} }
			\hline  
			\backslashbox{Alg.}{SNR (dB)}   &   0      &   5    & 10 & 15 & 20   \\
			\hline \hline 
			GMAP-AEM & 0.12 &	0.17 &	0.26 &	0.32 &	0.44   \\
			\hline
			
			BOX-AIEM &0.59 &	0.92 &	0.95 &	1.60 &	2.06  \\
			
			\hline 
			
			DIEM &  \multicolumn{5}{c}{0.082}\\
			\hline
		\end{tabular}
	}
\end{table}

\section{Conclusion}
\label{sect:conclusion}

To conclude, the EM convergence rate was analyzed for a class of approximate ML formulations for OMOD.
The analysis revealed the possibility of slow convergence under high SNRs.
It also led to accelerated and inexact EM schemes.
The implementation of these schemes for OMOD was studied,
and the use of deep unfolding in EM for OMOD was also considered.
The resulting EM algorithms were numerically shown to have competitive performance in the OMOD application.
 In deep learning-based methods, empirical study plays a key role. As a future direction, it would be useful to use numerical experiments to thoroughly examine the reliability and efficiency of the proposed deep unfolding EM method under different conditions.

\appendix
\section*{Appendix}

\section{Proof of Proposition~\ref{prop:Lf}}
\label{app:prop1}

We will need the following result.
\begin{Lemma} \label{lem:0}
	Consider the function $\psi(z)$ in \eqref{eq:psi_def}.
	The function $\psi(z)$ is $1$-Lipschitz continuous on $\Rbb$, i.e.,
	$| \psi(z) - \psi(z') | \leq 1$ for all $z,z' \in \Rbb$.
\end{Lemma}

{\em Proof of Lemma~\ref{lem:0}:}
A differentiable function $f: \Rbb \rightarrow \Rbb$ is $L$-Lipschitz continuous on $\Rbb$ if $| {\rm d} f(x)/{\rm d}x | \leq L$ for all $x$.
Moreover, $\psi(z)$ is closely related to Mill's ratio, and it was shown in the context therein that  ${\rm d} \psi(z)/{\rm d}z  \leq 1$ for all $z$ \cite{sampford1953some}.
The desired result follows.
\hfill $\blacksquare$
\medskip

From the expression of $\nabla f(\btheta)$ in \eqref{eq:ana_df}, we have
\begin{align*}
	\| \nabla f(\btheta) - \nabla f(\btheta') \| & = \| \bB^\top (\psi(\bB\btheta ) - \psi(\bB\btheta') ) \| \\
	& \leq \sigma_{\rm max}(\bB) \| \psi(\bB\btheta ) - \psi(\bB\btheta') \| \\
	& \leq \sigma_{\rm max}(\bB) \| \bB\btheta - \bB\btheta' \| \\
	& \leq \sigma_{\rm max}(\bB)^2 \| \btheta - \btheta' \|,
\end{align*}
where we apply Lemma~\ref{lem:0} to obtain the second inequality.
The proof is complete.

\ifconfver
\begin{figure*}[b]
	\normalsize
\setcounter{mytempeqncnt}{\value{equation}}
	\setcounter{equation}{77}
	\hrulefill
	\begin{subequations} \label{eq:lem:1:proof4}
		\begin{align}
		F(\btheta^{k+1}) & \leq g(\btheta^{k+1}|\btheta^k) + h(\btheta^{k+1})
		\label{eq:lem:1:proof4a} \\
		&
		= f(\btheta^k) +  \langle \nabla f(\btheta^k), \btheta^{k+1} - \btheta^k \rangle
		+ \tfrac{1}{2} \| \bB(\btheta^{k+1} - \btheta^k) \|^2  + h(\btheta^{k+1})
		\label{eq:lem:1:proof4b} \\
		&  \leq f(\tbtheta) + \langle \nabla f(\btheta^k), \btheta^{k+1} - \tbtheta \rangle + \tfrac{1}{2} \| \bB(\btheta^{k+1} - \btheta^k) \|^2  + h(\btheta^{k+1})
		\label{eq:lem:1:proof4c} \\
		&  \leq F(\tbtheta) + \langle \bB^\top \bB ( \btheta^{k+1} - \btheta^k) - \be^{k+1}, \tbtheta - \btheta^{k+1} \rangle + \tfrac{1}{2} \| \bB(\btheta^{k+1} - \btheta^k) \|^2
		\label{eq:lem:1:proof4d} \\
		&  \leq F(\tbtheta) + \eps_{k+1}\| \tbtheta - \btheta^{k+1} \| + \langle \bB^\top \bB ( \btheta^{k+1} - \btheta^k), \tbtheta - \btheta^{k+1} \rangle + \tfrac{1}{2} \| \bB(\btheta^{k+1} - \btheta^k) \|^2
		\label{eq:lem:1:proof4d2} \\
		&  = F(\tbtheta) + \eps_{k+1}\| \tbtheta - \btheta^{k+1} \| + \tfrac{1}{2} \| \bB(\btheta^k - \tbtheta) \|^2 - \tfrac{1}{2} \| \bB(\btheta^{k+1} - \tbtheta) \|^2,
		\label{eq:lem:1:proof4e}
		\end{align}
	\end{subequations}
	\setcounter{equation}{\value{mytempeqncnt}}
\end{figure*}
\fi


\section{Proof of Propositions~\ref{prop:EM} and \ref{prop:AEM}--\ref{prop:NIEM}
}

The proofs of Propositions~\ref{prop:EM}, \ref{prop:AEM}, \ref{prop:IEM}, \ref{prop:AIEM} and \ref{prop:NIEM} follow the same principles as their PG counterparts \cite{beck2017first,schmidt2011convergence}, although a careful verification remains necessary.
The following lemma, which considers the structure of our problem,
is important.

\begin{Lemma} \label{lem:1}
	Consider problem \eqref{eq:ana_P} and the accompanied problem setup.
	Let $\btheta^k$ be given, and consider the following problem
	\beq \label{eq:ana_IEM_4proof}
	\btheta^{k+1} \approx \arg \min_{\btheta \in \Rbb^n}  G(\btheta|\btheta^k):= g(\btheta|\btheta^k) + h(\btheta),
	\eeq
	where we consider the alternate form of 	$g(\btheta|\btheta')$ in \eqref{eq:ana_u_EM};
	and
	$\btheta^{k+1}$ denotes an approximate solution that satisfies
	\beq \label{eq:ana_IEM_eps_4proof}
	{\rm dist}(\bzero, \partial G(\btheta^{k+1}|\btheta^k)) \leq \eps_{k+1}.
	\eeq
	Let $\tilde{\btheta}$ be any point on $\Rbb^n$. It holds that
	\ifconfver
	\begin{align}
	F(\btheta^{k+1}) - F(\tbtheta) & \leq \tfrac{1}{2} \| \bB(\btheta^k - \tbtheta) \|^2 - \tfrac{1}{2} \| \bB(\btheta^{k+1} - \tbtheta) \|^2 \nonumber \\
	& + \eps_{k+1} \| 	\btheta^{k+1} - \tbtheta \|. \label{eq:lem:1}
	\end{align}
	\else
	\beq
	F(\btheta^{k+1}) - F(\tbtheta)  \leq \tfrac{1}{2} \| \bB(\btheta^k - \tbtheta) \|^2 - \tfrac{1}{2} \| \bB(\btheta^{k+1} - \tbtheta) \|^2  + \eps_{k+1} \| 	\btheta^{k+1} - \tbtheta \|.
	\label{eq:lem:1}
	\eeq
	\fi
\end{Lemma}

\noindent{\em Proof of Lemma~\ref{lem:1}:}
First we list some basic results.
Since $f$ and $h$ are convex, we have, respectively,
\begin{align}
	f(\tbtheta) & \geq f(\btheta^k) + \langle \nabla f(\btheta^k), \tbtheta - \btheta^k \rangle,
	\label{eq:lem:1:proof1}
	\\
	h(\tbtheta) & \geq h(\btheta^{k+1}) + \langle \bv^{k+1}, \tbtheta - \btheta^{k+1} \rangle,
	\label{eq:lem:1:proof2}
\end{align}
for any $\bv^{k+1} \in \partial h(\btheta^{k+1})$.
From \eqref{eq:ana_IEM_eps_4proof}, we have
\[
\be^{k+1} \in \nabla g(\btheta^{k+1}|\btheta^k)+ \partial h(\btheta^{k+1}),
\]
for some $\| \be^{k+1} \| \leq \eps_{k+1}$.
From \eqref{eq:ana_u_EM}, it can be shown that
\beq \label{eq:lem:1:proof_dg}
\nabla g(\btheta|\btheta') = \nabla f(\btheta') + \bB^\top \bB( \btheta - \btheta').
\eeq
The above two equations imply that
\beq
\be^{k+1} = \nabla f(\btheta^k) + \bB^\top \bB( \btheta^{k+1} - \btheta^k) + \bv^{k+1},
\label{eq:lem:1:proof3}
\eeq
for some $\bv^{k+1} \in \partial h(\btheta^{k+1})$, $\| \be^{k+1} \| \leq \eps_{k+1}$.

Second, we put together the above results to show \eqref{eq:lem:1}.
\ifconfver
We have the derivations in \eqref{eq:lem:1:proof4} at the bottom of this page,
\setcounter{equation}{78}
\else
We have
	\begin{subequations} \label{eq:lem:1:proof4}
	\begin{align}
	F(\btheta^{k+1}) & \leq g(\btheta^{k+1}|\btheta^k) + h(\btheta^{k+1})
	\label{eq:lem:1:proof4a} \\
	&
	= f(\btheta^k) +  \langle \nabla f(\btheta^k), \btheta^{k+1} - \btheta^k \rangle
	+ \tfrac{1}{2} \| \bB(\btheta^{k+1} - \btheta^k) \|^2  + h(\btheta^{k+1})
	\label{eq:lem:1:proof4b} \\
	&  \leq f(\tbtheta) + \langle \nabla f(\btheta^k), \btheta^{k+1} - \tbtheta \rangle + \tfrac{1}{2} \| \bB(\btheta^{k+1} - \btheta^k) \|^2  + h(\btheta^{k+1})
	\label{eq:lem:1:proof4c} \\
	&  \leq F(\tbtheta) + \langle \bB^\top \bB ( \btheta^{k+1} - \btheta^k) - \be^{k+1}, \tbtheta - \btheta^{k+1} \rangle + \tfrac{1}{2} \| \bB(\btheta^{k+1} - \btheta^k) \|^2
	\label{eq:lem:1:proof4d} \\
	&  \leq F(\tbtheta) + \eps_{k+1}\| \tbtheta - \btheta^{k+1} \| + \langle \bB^\top \bB ( \btheta^{k+1} - \btheta^k), \tbtheta - \btheta^{k+1} \rangle + \tfrac{1}{2} \| \bB(\btheta^{k+1} - \btheta^k) \|^2
	\label{eq:lem:1:proof4d2} \\
	&  = F(\tbtheta) + \eps_{k+1}\| \tbtheta - \btheta^{k+1} \| + \tfrac{1}{2} \| \bB(\btheta^k - \tbtheta) \|^2 - \tfrac{1}{2} \| \bB(\btheta^{k+1} - \tbtheta) \|^2,
	\label{eq:lem:1:proof4e}
	\end{align}
\end{subequations}
\fi
where \eqref{eq:lem:1:proof4a} is due to $f(\btheta) \leq g(\btheta|\btheta')$;
\eqref{eq:lem:1:proof4b} is obtained by putting \eqref{eq:ana_u_EM} into \eqref{eq:lem:1:proof4a};
\eqref{eq:lem:1:proof4c} is obtained by putting  \eqref{eq:lem:1:proof1} into \eqref{eq:lem:1:proof4b};
\eqref{eq:lem:1:proof4d} is obtained by putting \eqref{eq:lem:1:proof2} and \eqref{eq:lem:1:proof3} into \eqref{eq:lem:1:proof4c};
\eqref{eq:lem:1:proof4d2} is due to the Cauchy-Schwarz inequality and $\| \be^{k+1} \| \leq \eps_{k+1}$;
and \eqref{eq:lem:1:proof4e} is obtained by appropriate manipulation of the terms in \eqref{eq:lem:1:proof4d2}.
The proof is complete.
\hfill $\blacksquare$
\medskip

\subsection{Proof of Proposition~\ref{prop:EM}}
\label{app:prop2} ~\smallskip

We start with the basic EM method.
Applying Lemma~\ref{lem:1} to the EM method \eqref{eq:ana_EM}, with $\tbtheta = \btheta^\star$ and $\eps_k = 0$, we have
\[
F(\btheta^{k+1}) - F(\btheta^\star) \leq \tfrac{1}{2}  ( \| \bB(\btheta^k - \btheta^\star) \|^2 -  \| \bB(\btheta^{k+1} - \btheta^\star) \|^2 ),
\]
for  $k \geq 0$.
It follows that, for $k \geq 1$,
\[
\sum_{i=0}^{k-1} ( F(\btheta^{i+1}) - F(\btheta^\star))
\leq \tfrac{1}{2}  ( \| \bB(\btheta^0 - \btheta^\star) \|^2 -  \| \bB(\btheta^{k} - \btheta^\star) \|^2 ).
\]
Also, by the basic EM property $F(\btheta^{k+1}) \leq F(\btheta^k)$,
we have
\beq \label{eq:prop_EM_proof1}
\begin{aligned}
F(\btheta^{k}) - F(\btheta^\star) & \textstyle \leq \frac{1}{k} \sum_{i=0}^{k-1} ( F(\btheta^{i+1}) - F(\btheta^\star)) \\
& \leq \tfrac{1}{2k} \| \bB(\btheta^0 - \btheta^\star) \|^2.
\end{aligned}
\nonumber
\eeq

\subsection{Proof of Proposition~\ref{prop:IEM}}
\label{app:prop5} ~\smallskip

Next, we consider the inexact EM scenario in \eqref{eq:ana_IEM}--\eqref{eq:ana_IEM_desc}.
For convenience, let $w_k = \sum_{i=0}^{k-1} ( F(\btheta^{i+1}) - F(\btheta^\star))$.
Following the same proof in the preceding subsection, but with $\eps_k \neq 0$ in general, one can show that, for $k \geq 1$,
\beq \label{eq:prop_IEM_proof1}
w_k + \tfrac{1}{2} \| \bB(\btheta^{k} - \btheta^\star) \|^2 \leq \tfrac{1}{2}  \| \bB(\btheta^0 - \btheta^\star) \|^2  + \sum_{i=1}^k \eps_i \| \btheta^i - \btheta^\star \|,
\eeq
\beq \label{eq:prop_IEM_proof0}
F(\btheta^{k}) - F(\btheta^\star) \leq \tfrac{1}{k} w_k.
\eeq
Note that \eqref{eq:prop_IEM_proof0} is due to the descent condition $F(\btheta^{k+1}) \leq F(\btheta^k)$, enabled by the condition $G(\btheta^{k+1}|\btheta^k) \leq G(\btheta^{k}|\btheta^k)$ in \eqref{eq:ana_IEM_desc}.
The new challenge lies in the last term of \eqref{eq:prop_IEM_proof1}.
To deal with it, assume that $\bB$ has full column rank.
We have
\beq \label{eq:prop_IEM_proof2}
\textstyle
\sum_{i=1}^k \eps_i \| \btheta^i - \btheta^\star \| \leq \frac{1}{\sigma_{\rm min}(\bB)} \sum_{i=1}^k \eps_i \| \bB(\btheta^i - \btheta^\star) \|,
\eeq
where we have used the result $\| \bB \bx \| \geq \sigma_{\rm min}(\bB) \| \bx \|$.
Since $w_k \geq 0$, Eqs.~\eqref{eq:prop_IEM_proof1} and \eqref{eq:prop_IEM_proof2} imply that
\beq \label{eq:prop_IEM_proof3}
\tfrac{1}{2} \| \bB(\btheta^{k} - \btheta^\star) \|^2
\leq
\tfrac{1}{2} \| \bB(\btheta^0 - \btheta^\star) \|^2 + \sum_{i=1}^k
\frac{\eps_i}{\sigma_{\rm min}} \| \bB(\btheta^i - \btheta^\star) \|,
\eeq
where we denote $\sigma_{\rm min} = \sigma_{\rm min}(\bB)$ for convenience.
Consider the following lemma.
\begin{Lemma}\cite[Lemma 1]{schmidt2011convergence} \label{lem:iem_seq}
	Let $\{ u_k \}_{k \geq 0}$ be a non-negative sequence. Suppose that, for $k \geq 1$,
	\[ \textstyle
	u_k^2 \leq u_0^2 + \sum_{i=1}^k \lambda_i u_i,
	\]
	where $\lambda_i \geq 0$ for all $i$.
	Then, for $k \geq 1$, we have
	\[
	u_k \leq \textstyle  \tfrac{1}{2} \sum_{i=1}^k \lambda_i + \left( u_0^2 + \left( \tfrac{1}{2} \sum_{i=1}^k \lambda_i \right)^2 \right)^{1/2}.
	\]
\end{Lemma}
Applying Lemma~\ref{lem:iem_seq} to \eqref{eq:prop_IEM_proof3} gives
\beq \label{eq:prop_IEM_proof4}
\tfrac{1}{\sqrt{2}} \| \bB(\btheta^{k} - \btheta^\star) \| \leq A_k + (u_0^2 + A_k^2 )^{1/2} \leq u_0 + 2A_k,
\eeq
for $k \geq 1$,
where
\beq \label{eq:prop_IEM_proof5}
u_0 = \tfrac{1}{\sqrt{2}} \| \bB(\btheta^{0} - \btheta^\star) \|, \quad A_k = \textstyle \tfrac{1}{\sqrt{2} \sigma_{\rm min}} \sum_{i=1}^k \eps_i.
\eeq

Now we use \eqref{eq:prop_IEM_proof2} and \eqref{eq:prop_IEM_proof4} to bound the last term of \eqref{eq:prop_IEM_proof1}.
From  \eqref{eq:prop_IEM_proof1}, we have

\begin{align}
w_k & \leq \textstyle  \tfrac{1}{2} \| \bB(\btheta^0 - \btheta^\star ) \|^2 + \sum_{i=1}^k \eps_i \| \btheta^i - \btheta^\star \| \nonumber \\
& \leq \textstyle
u_0^2
+ \sum_{i=1}^k \frac{\sqrt{2} \eps_i}{\sigma_{\rm min}} (u_0 + 2 A_i) \nonumber \\
& \textstyle  \leq u_0^2 + \sum_{i=1}^k \frac{\sqrt{2} \eps_i}{\sigma_{\rm min}} (u_0 + 2 A_k ) = u_0^2 + 2 A_k ( u_0 + 2 A_k ) \nonumber  \\
& = (u_0 + 2 A_k )^2,
\label{eq:prop_IEM_proof6}
\end{align}
where the second inequality is obtained by applying \eqref{eq:prop_IEM_proof2} and \eqref{eq:prop_IEM_proof4};
and the third inequality is due to $A_{k} \geq A_i$ for $i \leq k$.
Applying \eqref{eq:prop_IEM_proof6} and \eqref{eq:prop_IEM_proof5} to \eqref{eq:prop_IEM_proof0} leads to the final result in Proposition~\ref{prop:IEM}.

\subsection{Proof of Proposition~\ref{prop:AEM}}
\label{app:prop4} ~\smallskip

To describe the proof for the accelerated EM method, we first write down some basic results \cite{beck2009fast}.
	The sequence $\{ t_k \}_{k \geq 0}$ in \eqref{eq:ana_FISTA2} satisfies
	(a) $t_k^2 - t_k = t_{k+1}^2$; (b) $t_k \geq (k+2)/2$; and (c) $t_k \leq k+1$.
Also, the following result is important.
\begin{Lemma} \label{lem:aem_1}
	Let
	$\tbtheta = t_k^{-1} \btheta^\star + (1- t_k^{-1} ) \btheta^k$.
	It holds that
	\beq \label{eq:aem_proof1}
	t_k^2 (F(\btheta^{k+1}) - F(\tbtheta)) \geq t_k^2 v_{k+1} - t_{k-1}^2 v_k,
	\eeq
	where $v_k = F(\btheta^k) - F(\btheta^\star)$.
\end{Lemma}
Note that Lemma~\ref{lem:aem_1} uses only the convexity of $F$ and the property $t_k^2 - t_k = t_{k+1}^2$.
Applying Lemma~\ref{lem:1} to the accelerated EM method \eqref{eq:ana_AEM}--\eqref{eq:ana_FISTA2}, with $\tbtheta$ given by the one in Lemma~\ref{lem:aem_1}, $\btheta^k= \bvartheta^k$  and $\eps_k = 0$, we have
\beq \label{eq:aem_proof2}
F(\btheta^{k+1}) - F(\btheta^\star) \leq \tfrac{1}{2}  ( \| \bB(\bvartheta^k - \tbtheta) \|^2 -  \| \bB(\btheta^{k+1} - \tbtheta) \|^2 ),
\eeq
for $k \geq 0$.
Let
$\bu^{k+1} = t_k \btheta^{k+1} - ( (t_k - 1 ) \btheta^k + \btheta^\star)$ for $k \geq 0$.
It can be verified that, for $k \geq 1$,
\begin{align}
\btheta^{k+1} - \tbtheta & = t_k^{-1} \bu^{k+1},
\quad
\bvartheta^k -  \tbtheta
= t_k^{-1} \bu^{k}. \label{eq:aem_proof4}
\end{align}
Combining \eqref{eq:aem_proof1}--\eqref{eq:aem_proof4} gives
\beq \label{eq:aem_proof5}
t_k^2 v_{k+1} +  \tfrac{1}{2}  \| \bB \bu^{k+1} \|^2  \leq t_{k-1}^2 v_{k} +  \tfrac{1}{2}  \| \bB \bu^{k} \|^2, \quad k \geq 1.
\nonumber
\eeq
The above inequality further implies that
\beq \label{eq:aem_proof6}
t_k^2 v_{k+1} +  \tfrac{1}{2}  \| \bB \bu^{k+1} \|^2 \leq t_0^2 v_1 + \tfrac{1}{2}  \| \bB \bu^{1} \|^2.
\eeq
Applying Lemma~\ref{lem:1} to the accelerated EM method \eqref{eq:ana_AEM}, with $k=1$, $\tbtheta = \btheta^\star$  and $\eps_k = 0$, and noting $\bvartheta^0 = \btheta^0$, we have
\beq
v_1 \leq \tfrac{1}{2} \| \bB ( \btheta^0 - \btheta^\star ) \|^2 - \tfrac{1}{2} \| \bB ( \btheta^1 - \btheta^\star ) \|^2.
\nonumber
\eeq
By noting that $t_0 = 1$ and $\bu^1 = \btheta^1 - \btheta^\star$, we further obtain
\beq
t_0^2 v_1 + \tfrac{1}{2}  \| \bB \bu^{1} \|^2 \leq \tfrac{1}{2} \| \bB ( \btheta^0 - \btheta^\star ) \|^2.
\nonumber
\eeq
Applying the above inequality to \eqref{eq:aem_proof6} gives
\beq
t_k^2 v_{k+1} \leq \tfrac{1}{2} \| \bB ( \btheta^0 - \btheta^\star ) \|^2.
\nonumber
\eeq
Finally, by applying $t_k \geq (k+2)/2$ to the above inequality, we obtain the final result in Proposition~\ref{prop:AEM}.

\subsection{Proof of Proposition~\ref{prop:AIEM}}
\label{app:prop6} ~\smallskip

The proof for the accelerated inexact EM scenario is a combination of the proof methods in the last two subsections.
Following the same proof in the previous subsection, with $\eps_k \neq 0$ in general, one can show that, for $k \geq 1$,
\beq \label{eq:aiem_proof1}
t_k^2 v_{k+1} +  \tfrac{1}{2}  \| \bB \bu^{k+1} \|^2  \leq t_{k-1}^2 v_{k} +  \tfrac{1}{2}  \| \bB \bu^{k} \|^2 + \tfrac{t_k \eps_{k+1}}{\sigma_{\rm min}} \| \bB \bu^{k+1} \|,
\nonumber
\eeq
and that
\beq
t_0^2 v_1 + \tfrac{1}{2}  \| \bB \bu^{1} \|^2  \leq \tfrac{1}{2} \| \bB ( \btheta^0 - \btheta^\star ) \|^2 + \tfrac{t_0 \eps_{1}}{\sigma_{\rm min}} \| \bB \bu^1 \|.
\nonumber
\eeq
The above two inequalities imply that, for $k \geq 0$,
\beq  \label{eq:aiem_proof2}
t_k^2 v_{k+1} +  \tfrac{1}{2}  \| \bB \bu^{k+1} \|^2 \leq u_0^2 + \sum_{i=1}^{k+1} \frac{t_{i-1} \eps_i}{\sigma_{\rm min}} \| \bB \bu^i \|,
\eeq
where $u_0 =  \| \bB ( \btheta^0 - \btheta^\star ) \| / \sqrt{2}$.
Following the same proof in the last last subsection (specifically, by seeing  \eqref{eq:prop_IEM_proof3} as \eqref{eq:aiem_proof2}),
one can show that
\beq \label{eq:aiem_proof4}
t_k^2 v_{k+1} \leq ( u_0 + 2 A_{k+1} )^2,
\nonumber
\eeq
where $A_{k+1} = ( \sum_{i=1}^{k+1} t_{i-1} \eps_i )/ (\sqrt{2} \sigma_{\rm min})$.
Finally, by applying $t_k \geq (k+2)/2$ and $t_k \leq k+1$ to the left-hand and right-hand sides of the above equation, respectively, we obtain the final result in Proposition~\ref{prop:AIEM}.

\subsection{Proof of Proposition~\ref{prop:NIEM}}
\label{app:prop7} ~\smallskip

The proof for the inexact EM scenario \eqref{eq:ana_IEM}--\eqref{eq:mod_G} in the case of non-convex $h$ is as follows.
First, we show that
\beq  \label{eq:niem_proof_core1}
\min_{i=0,\ldots,k-1} \tfrac{2}{\tau} \| \bB(\btheta^{i+1}-\btheta^i) \|^2 + \delta_{i+1}
\leq \frac{1}{k} \left( C_1 + \sum_{i=1}^k \delta_i \right)
\eeq
for $k \geq 1$ and for any
$\delta_1,\ldots,\delta_k \geq 0$,
where $C_1 = F(\btheta^0) - F(\btheta^\star)$.
From the expression of $G(\btheta|\btheta')$ in \eqref{eq:mod_G},
we have
\begin{align*}
F(\btheta^{i+1}) + \tfrac{\tau}{2} \| \bB(\btheta^{i+1}-\btheta^i) \|^2 &
\leq G(\btheta^{i+1}|\btheta^i)  \leq G(\btheta^{i}|\btheta^i)  \\
& = F(\btheta^i),
\end{align*}
where the first inequality is due to $f(\btheta) \leq g(\btheta|\btheta')$, and the second inequality is due to the descent condition \eqref{eq:ana_IEM_desc}.
Summing the above equation over $i=0,\ldots,k-1$ yields
\[ \textstyle
\tfrac{\tau}{2} \sum_{i=0}^{k-1} \| \bB(\btheta^{i+1}-\btheta^i) \|^2
\leq C_1,
\]
and further,
\[ \textstyle
 \sum_{i=0}^{k-1} ( \tfrac{\tau}{2} \| \bB(\btheta^{i+1}-\btheta^i) \|^2 +\delta_{i+1})
\leq C_1 + \sum_{i=0}^{k-1} \delta_{i+1}.
\]
Applying $\sum_{i=0}^{k-1} u_i \geq k \min_{i=0,\ldots,k-1} u_i$ to the left-hand side of the above equation leads to the result in \eqref{eq:niem_proof_core1}.

Second, we show that
\beq \label{eq:niem_proof_core2}
{\rm dist}_\bR(\bzero,\partial F(\btheta^{k+1})) \leq (2 + \tau) \| \bB(\btheta^{k+1}-\btheta^k) \| + \frac{\eps_{k+1}}{\sigma_{\rm min}},
\eeq
where $\sigma_{\rm min} = \sigma_{\rm min}(\bB)$.
From the $\eps$-criticality condition \eqref{eq:ana_IEM_eps}, the structure of $G(\btheta|\btheta')$ in \eqref{eq:mod_G}, and the expressions of  $g(\btheta|\btheta')$ in \eqref{eq:lem:1:proof_dg}, it can be shown that
\[
\be^{k+1} = \nabla f(\btheta^k) + ( 1 + \tau ) \bB^\top \bB (\btheta^{k+1} - \btheta^k) + \bv^{k+1},
\]
for some $\bv^{k+1} \in \partial h(\btheta^{k+1})$ and $\| \be^{k+1} \| \leq \eps_{k+1}$.
This, together with $\bR = (\bB^\top \bB)^{-1}$, give rise to
\ifconfver
\begin{align*}
& {\rm dist}_\bR(\bzero,\partial F(\btheta^{k+1}))  = {\rm dist}_\bR(\bzero, \nabla f(\btheta^{k+1}) + \partial h(\btheta^{k+1})) \\
&  \leq \| \nabla f(\btheta^{k+1}) - \nabla f(\btheta^{k}) -
(1+\tau) \bR^{-1} (\btheta^{k+1} - \btheta^k) + \be^{k+1} \|_\bR \\
&  \leq \| \nabla f(\btheta^{k+1}) - \nabla f(\btheta^{k})  \|_\bR +
(1 + \tau) \| \bR^{-1} (\btheta^{k+1} - \btheta^k) \|_\bR  \\
& ~~ + \tfrac{\eps_{k+1}}{\sigma_{\rm min}},
\end{align*}
\else
\begin{align*}
 {\rm dist}_\bR(\bzero,\partial F(\btheta^{k+1})) &  = {\rm dist}_\bR(\bzero, \nabla f(\btheta^{k+1}) + \partial h(\btheta^{k+1})) \\
&  \leq \| \nabla f(\btheta^{k+1}) - \nabla f(\btheta^{k}) -
(1+\tau) \bR^{-1} (\btheta^{k+1} - \btheta^k) + \be^{k+1} \|_\bR \\
&  \leq \| \nabla f(\btheta^{k+1}) - \nabla f(\btheta^{k})  \|_\bR +
(1 + \tau) \| \bR^{-1} (\btheta^{k+1} - \btheta^k) \|_\bR   + \tfrac{\eps_{k+1}}{\sigma_{\rm min}},
\end{align*}
\fi
where we have used the fact that $\| \bx \|_\bR \leq \sigma_{\rm max}(\bR)^{1/2} \| \bx \|$ and $\sigma_{\rm max}(\bR)^{1/2} = 1/\sigma_{\rm min}(\bB)$.
It can be verified that
\begin{align*}
\| \bR^{-1} (\btheta^{k+1} - \btheta^k) \|_\bR & = \| \bB (\btheta^{k+1} - \btheta^k) \|.
\end{align*}
Let $\bu = \psi(\bB\btheta^k) - \psi(\bB\btheta^{k+1})$ for convenience.
We have
\begin{align*}
	\| \nabla f(\btheta^{k+1}) - \nabla f(\btheta^{k})  \|_\bR
& = \bu^\top \bB(\bB^\top \bB)^{-1} \bB^\top \bu \\
&   \leq \| \bu \|^2 = \| \psi(\bB\btheta^k) - \psi(\bB\btheta^{k+1}) \|^2 \\
&  \leq \| \bB\btheta^k - \bB\btheta^{k+1} \|^2,
\end{align*}
where the first inequality is due to the fact that $\bB(\bB^\top \bB)^{-1} \bB^\top$ is an orthogonal projection matrix, and the second inequality is due to Lemma~\ref{lem:0}.
Putting the above components together, we have \eqref{eq:niem_proof_core2}.

Third, we use \eqref{eq:niem_proof_core1} and \eqref{eq:niem_proof_core2} to show the desired result.
From \eqref{eq:niem_proof_core2}, we have
\beq
{\rm dist}_\bR(\bzero,\partial F(\btheta^{i+1}))^2 \leq 2 \left( (2 + \tau)^2 \| \bB(\btheta^{i+1}-\btheta^i) \|^2 + \tfrac{\eps_{i+1}^2}{\sigma_{\rm min}^2} \right),
\nonumber
\eeq
which is due to $(x+y)^2 \leq 2 x^2 + 2 y^2$.
By appropriately setting the $\delta_i$'s in \eqref{eq:niem_proof_core1}, we have
\beq
\min_{i=0,\ldots,k-1} \tfrac{\tau}{2} \| \bB(\btheta^{i+1}-\btheta^i) \|^2 + \tfrac{\tfrac{\tau}{2} \eps_{k+1}^2}{(2+\tau)^2 \sigma_{\rm min}^2}
\leq \frac{1}{k} \left( C_1 + \tfrac{ \tfrac{\tau}{2} C_2}{(2+\tau)^2} \right),
\nonumber
\eeq
where $C_2 = \sum_{i=1}^k \eps_{i}^2/ \sigma_{\rm min}^2$.
Combining the above two equations gives
\beq
\min_{i=0,\ldots,k-1} {\rm dist}_\bR(\bzero,\partial F(\btheta^{i+1}))^2
\leq \frac{4}{k \tau} ( (2+\tau)^2 C_1  + \tfrac{\tau}{2} C_2),
\nonumber
\eeq
which leads to the final result of Proposition~\ref{prop:NIEM}.

\section{Proof of Proposition~\ref{prop:new_EM_PG}}
\label{app:prop3}

\subsection{The Case with PG, $h(\btheta)= \Ind_\setX(\btheta)$} ~\smallskip

Suppose that $\eps \geq \| \btheta^\star - \btheta^k \|$. By the triangle inequality,
\begin{subequations} \label{eq:k_bnd_nuproof1_00}
\begin{align}
	\eps & \geq \| \btheta^\star - \btheta^k \|  \geq \| \btheta^\star - \btheta^0 \| - \| \btheta^0 - \btheta^k \|,
	\label{eq:k_bnd_nuproof1_0} \\
	& \geq
	\| \btheta^\star - \btheta^0 \| - \textstyle \sum_{i=0}^{k-1} \| \btheta^{i} - \btheta^{i+1} \|.
	\label{eq:k_bnd_nuproof1_1}
\end{align}
\end{subequations}
We focus on $\| \btheta^{i} - \btheta^{i+1} \|$.
Consider the following result.
\begin{Lemma}\cite[Theorem 6.42]{beck2017first} \label{lem:prox2}
	For any convex $h: \Rbb^n \rightarrow \Rbb \cup \{ +\infty \}$, we have $\| \prox_h(\bz_1) - \prox_h(\bz_2) \| \leq \| \bz_1 - \bz_2 \|$ for any $\bz_1, \bz_2$.
\end{Lemma}
Also, for $h(\btheta)= \Ind_\setX(\btheta)$, $\setX$ being non-empty closed convex, 
we have
$\prox_{\eta h}(\bz) = \arg \min_{\btheta \in \setX} \| \bz - \btheta \|^2.$
From the above equation, it can be verified that the PG iterates in \eqref{eq:ana_PG} satisfy
\beq
\btheta^i = \prox_{\eta h}(\btheta^i). \nonumber
\eeq
Applying the above results to the PG method \eqref{eq:ana_PG}, we have
\begin{align}
	\| \btheta^{i} - \btheta^{i+1} \| & = \| \prox_{\eta h}(\btheta^{i}) - \prox_{\eta h}(\btheta^{i} - \eta \nabla f(\btheta^i)) \| \nonumber \\
	& \leq \| \eta \nabla f(\btheta^i) \|.
	\label{eq:k_bnd_nuproof1_3}
\end{align}
Also, from the expression of $\nabla f(\btheta)$ in \eqref{eq:ana_df}, we have
\begin{align}
	\| \nabla f(\btheta) \| & = \| \bB^\top \psi(\bB\btheta)) \|
	\leq \sigma_1 \| \psi(\bB\btheta) \|  \leq \sigma_1 \sqrt{m},	
	\label{eq:k_bnd_nuproof1_4}
\end{align}
where we denote $\sigma_1 = \sigma_{\rm max}(\bB)$ for convenience, and
the second inequality is due to $0 < \psi(z) < 1$ for all $z$ (see \eqref{eq:psi_def}).
By putting \eqref{eq:k_bnd_nuproof1_4} into \eqref{eq:k_bnd_nuproof1_3}, and then \eqref{eq:k_bnd_nuproof1_3} into \eqref{eq:k_bnd_nuproof1_1}, and by setting $\eta = 1/\sigma_1^2$, we obtain
$\eps \geq \| \btheta^\star - \btheta^0 \| - \sigma_1^{-1} \sqrt{m} k,$
and consequently
the desired result
\[
k \geq
\sigma_1 (\sqrt{m})^{-1}
( \| \btheta^\star - \btheta^0 \| -\eps ).
\]

\subsection{The Case with PG, $h(\btheta)= \lambda \| \btheta \|^2/2$} ~\smallskip

We begin with $\| \btheta^k - \btheta^0 \|$.
For $h(\btheta) = \lambda \| \btheta \|^2 /2$, $\lambda > 0$, it can be verified that $\prox_{\eta h}(\bz ) = \alpha^{-1} \bz$, where $\alpha = 1 + \eta \lambda$.
Applying this result to the PG method \eqref{eq:ana_PG} leads to
\begin{align*}
\btheta^k
& = \alpha^{-1} \btheta^{k-1} - \alpha^{-1} \eta \nabla f(\btheta^{k-1}) \\
& = \alpha^{-k} \btheta^0  - \eta \textstyle \sum_{i=0}^{k-1} \alpha^{-(k-i)} \nabla f(\btheta^i),
\end{align*}
and subsequently
\begin{align}
	\| \btheta^k - \btheta^0 \| & \leq  ( 1 - \alpha^{-k} ) \| \btheta^0 \| + \eta (\textstyle \sum_{i=0}^{k-1} \alpha^{-(k-i)}) \| \nabla f(\btheta^i) \| \nonumber \\
	& \leq ( 1 - \alpha^{-k} ) \| \btheta^0 \| + \eta \sigma_1 \sqrt{m} (\textstyle \sum_{i=0}^{k-1} \alpha^{-(k-i)}),
	\nonumber \\
	& \leq ( 1 - \alpha^{-k} ) \| \btheta^0 \| + \eta \sigma_1 \sqrt{m} \frac{\alpha^{-1} - \alpha^{-k-1}}{1-\alpha^{-1}}, \nonumber \\
	& = ( 1 - \alpha^{-k} ) ( \| \btheta^0 \| + \sigma_1 \sqrt{m} \lambda^{-1} ),
	\label{eq:k_bnd_nuproof2_3}
\end{align}
where the second inequality is due to \eqref{eq:k_bnd_nuproof1_4},
and the last equation is obtained by putting $\alpha = 1 + \eta \lambda$ into the equation.

Next, we apply \eqref{eq:k_bnd_nuproof2_3} to \eqref{eq:k_bnd_nuproof1_0} to obtain
\beq \label{eq:k_bnd_nuproof2_4}
1 - \alpha^{-k} \geq \frac{ \| \btheta^\star - \btheta^0 \| - \eps}{ \| \btheta^0 \| + 	\sigma_1 \sqrt{m} \lambda^{-1} } := C.
\eeq
Assume $C < 1$. Then \eqref{eq:k_bnd_nuproof2_4} can be rewritten as
\beq \label{eq:k_bnd_nuproof2_5}
\log(1-C)  \geq \log(\alpha^{-k}) = - k \cdot \log(\alpha).
\eeq
Applying $\log(z) \leq z - 1$ for $z > 0$ to both sides of \eqref{eq:k_bnd_nuproof2_5} gives
\beq \label{eq:k_bnd_nuproof2_6}
-C \geq -k (\alpha-1) = - k \eta \lambda
\eeq
By setting $\eta = 1/\sigma_1^{2}$ and putting
the expression of
$C$ in \eqref{eq:k_bnd_nuproof2_4}  into \eqref{eq:k_bnd_nuproof2_6}, we obtain the desired result
\beq \label{eq:k_bnd_nuproof2_7}
k \geq \frac{C}{\eta\lambda} =  \sigma_1^{2} \frac{ \| \btheta^\star - \btheta^0 \| - \eps}{ \lambda \| \btheta^0 \| + \sigma_1 \sqrt{m} }.
\nonumber
\eeq
What remains to show is that $C < 1$ is true for any $\lambda > 0$.
Since $\nabla F(\btheta^\star) = \bzero$, and $\nabla F(\btheta) = \nabla f(\btheta) + \lambda \btheta$, we have
\[
\| \btheta^\star \|
=  \lambda^{-1} \| \nabla f(\btheta^\star) \|  \leq \lambda^{-1} \sigma_1 \sqrt{m};
\]
the inequality is due to \eqref{eq:k_bnd_nuproof1_4}.
Consequently,
\beq  \label{eq:k_bnd_nuproof2_8}
\| \btheta^\star - \btheta^0 \| \leq \| \btheta^0 \| + \sigma_1 \sqrt{m}  \lambda^{-1},
\eeq
and it is seen that the $C$ in \eqref{eq:k_bnd_nuproof2_4} satisfies  $C< 1$ for any $\eps > 0$.
The proof is complete.

\subsection{The Case with EM, $h(\btheta)= \Ind_\setX(\btheta)$} ~\smallskip

To facilitate the proof, let
\beq \label{eq:k_bnd_nuproof3_1}
\bB = \bU_1 \tilde{\bSig} \bV_1^\top
\eeq
be the thin singular value decomposition of $\bB$,
where $\bU_1 \in \Rbb^{m \times r}$ and $\bV_1 \in \Rbb^{n \times r}$ are semi-orthogonal;
$r = {\rm rank}(\bB)$;
$\tilde{\bSig} = \Diag(\sigma_1,\ldots,\sigma_r)$;
$\sigma_1 \geq \cdots \geq \sigma_r > 0$.
First we show that the EM method \eqref{eq:ana_EM} can be expressed as
\beq \label{eq:k_bnd_nuproof3_2}
\btheta^{k+1} \in \Prox_{h,\bB}(\btheta^k +  \bW^\dag \bB^\top \psi(\bB\btheta^k) ),
\eeq
where $\bW = \bB^\top \bB$, and
\beq \label{eq:k_bnd_nuproof3_2a}
\Prox_{h,\bB}(\bz) = \arg \min_{\btheta} \tfrac{1}{2} \| \bB ( \bz - \btheta )  \|^2 + h(\btheta).
\eeq
To prove \eqref{eq:k_bnd_nuproof3_2}, we put $\nabla f(\btheta)$ in \eqref{eq:ana_df} into $g(\btheta|\btheta')$ in \eqref{eq:ana_u_EM}:
\begin{align}
	g(\btheta|\btheta') & =
		\tfrac{1}{2} \| \bB \btheta \|^2 - \langle \bW \btheta' + \bB^\top \psi(\bB\btheta'), \btheta \rangle + {\rm const.} \nonumber \\
		& = \tfrac{1}{2} \| \bB \btheta \|^2 - \langle \bW \btheta' + \bW \bW^\dag \bB^\top \psi(\bB\btheta'), \btheta \rangle + {\rm const.} \nonumber \\
		& = \tfrac{1}{2} \| \bB (\btheta - (\btheta'+ \bW^\dag \bB^\top \psi(\bB\btheta'))) \|^2 + {\rm const.}, \nonumber
\end{align}
where the second inequality is due to $\bB^\top = \bW \bW^\dag \bB^\top$, which can be verified to be true from \eqref{eq:k_bnd_nuproof3_1}.
The proof of \eqref{eq:k_bnd_nuproof3_2} is done.

Second, in a similar way as \eqref{eq:k_bnd_nuproof1_00}, we have
\begin{subequations}
\begin{align}
\eps & \geq \| \btheta^\star - \btheta^k \| \geq \| \bV_1^\top (\btheta^\star - \btheta^k) \|   \nonumber \\
& \geq \| \bV_1^\top(\btheta^\star - \btheta^0) \| - \| \bV_1^\top (\btheta^0 - \btheta^k ) \|
\label{eq:k_bnd_nuproof3_3a} \\
& \geq \| \bV_1^\top(\btheta^\star - \btheta^0) \| - \textstyle \sum_{i=0}^{k-1} \| \bV_1^\top(\btheta^{i} - \btheta^{i+1}) \|,
\label{eq:k_bnd_nuproof3_3}
\end{align}
\end{subequations}
where we have used the fact that $\| \bx \| \geq \| \bV_1^\top \bx \|$
for any semi-orthogonal $\bV_1$.
For $h(\btheta)= \Ind_\setX(\btheta)$, $\setX$ being non-empty closed convex, it can be verified that
$\Prox_{h, \bB}(\bz) = \arg \min_{\btheta \in \setX} \| \bB(\bz - \btheta) \|^2$,
and then, that
$\btheta^i \in \Prox_{h,\bB}(\btheta^i)$.
Consider an extension of Lemma~\ref{lem:prox2}.
\begin{Lemma} \label{lem:prox3}
	Let $h: \Rbb^n \rightarrow \Rbb \cup \{ + \infty \}$ be convex, let $\bz_1, \bz_2 \in \Rbb^n$, and let $\btheta_1 \in \Prox_{h,\bB}(\bz_1), \btheta_2 \in \Prox_{h,\bB}(\bz_2)$. We have
	$\| \bB( \btheta_1 - \btheta_2 ) \| \leq  \| \bB(\bz_1 - \bz_2) \|$.
\end{Lemma}

{\em Proof of Lemma~\ref{lem:prox3}}:
The proof is similar to that of Lemma~\ref{lem:prox2}; see, e.g., \cite{beck2017first}.
By the optimality condition of the problem in \eqref{eq:k_bnd_nuproof3_2a}, $\btheta_1$ and $\btheta_2$ must satisfy
\beq \label{eq:proof:lem:prox3_1}
\bB^\top \bB(\bz_1 - \btheta_1) \in \partial h(\btheta_1), ~ \bB^\top \bB(\bz_2 - \btheta_2) \in \partial h(\btheta_2).
\eeq
Since $h$ is convex, we have the following property
\beq \label{eq:proof:lem:prox3_2}
\langle \bg_1 - \bg_2, \btheta_1 - \btheta_2 \rangle \geq 0,
~ \text{for any $\bg_1 \in \partial h(\btheta_1), \bg_2 \in \partial h(\btheta_2)$;} \nonumber
\eeq
see, e.g., \cite{boyd2004convex}.
Applying \eqref{eq:proof:lem:prox3_1} to the above equation, we obtain
\begin{align*}
	& \langle \bB^\top \bB (\bz_1 - \btheta_1) - \bB^\top \bB (\bz_2 - \btheta_2), \btheta_1 - \btheta_2 \rangle \geq 0  \\
	\Longleftrightarrow ~ & \langle \bB^\top \bB(\bz_1 - \bz_2), \btheta_1 - \btheta_2 \rangle \geq \| \bB ( \btheta_1 - \btheta_2 ) \|^2  \\
	\Longrightarrow ~ & \| \bB (\bz_1 - \bz_2 ) \| \| \bB ( \btheta_1 - \btheta_2 ) \| \geq  \| \bB ( \btheta_1 - \btheta_2 ) \|^2 \\
	\Longleftrightarrow ~ &  \| \bB (\bz_1 - \bz_2 ) \| \geq \| \bB ( \btheta_1 - \btheta_2 ) \|,
\end{align*}
where the Cauchy-Schwarz inequality is used to obtain the third equation.
The proof is complete.$\blacksquare$

\medskip

Applying the above results to the equivalent form of EM in \eqref{eq:k_bnd_nuproof3_2}, we obtain
\begin{align}
	\| \bB(\btheta^{i} - \btheta^{i+1}) \| &
	\leq \| \bB (  \bW^\dag \bB^\top \psi(\bB\btheta^i) ) \|
	\leq \sqrt{m},
	\label{eq:k_bnd_nuproof3_4}
\end{align}
where we have also used $\sigma_{\rm max}(\bB  \bW^\dag \bB^\top ) = 1$, which can be verified from \eqref{eq:k_bnd_nuproof3_1},
and $\| \psi(\bB\btheta) \| \leq \sqrt{m}$ (see \eqref{eq:k_bnd_nuproof1_4}).
Applying \eqref{eq:k_bnd_nuproof3_4} to \eqref{eq:k_bnd_nuproof3_3}, and using $\| \bB \bx \| \geq \sigma_r \| \bV_1^\top \bx \|$, we obtain $\eps \geq \| \bV_1^\top (\btheta^\star - \btheta^0 ) \| - k \sigma_r^{-1} \sqrt{m}$, and consequently
\beq  \label{eq:k_bnd_nuproof3_5}
k \geq \sigma_r (\sqrt{m})^{-1} \| \bV_1^\top (\btheta^\star - \btheta^0 ) \|.
\eeq

Finally, it should be noted that $\sigma_r$ is the smallest positive singular value of $\bB$; and that $\| \bV_1^\top \bx \| = \| \bB^\top (\bB \bB^\top )^\dag \bB \bx \|$, which can be verified from \eqref{eq:k_bnd_nuproof3_1}.
Applying the above results to \eqref{eq:k_bnd_nuproof3_5} leads to the desired result.

\subsection{The Case with EM, $h(\btheta)= \lambda \| \btheta \|^2/2$} ~\smallskip

Our analysis starts with $\| \bV_1^\top (\btheta^0 - \btheta^k ) \|$.
For $h(\btheta) = \lambda \| \btheta \|^2 /2$, $\lambda > 0$, it can be shown from \eqref{eq:k_bnd_nuproof3_2}--\eqref{eq:k_bnd_nuproof3_2a} that
\begin{align*}
	\btheta^k & = \bPhi ( \btheta^{k-1} + \bW^\dag \bB^\top \psi(\bB\btheta^{k-1})) \\
	& = \bPhi^k \btheta^0 + \textstyle \sum_{i=0}^{k-1} \bPhi^{k-i} ( \bW^\dag \bB^\top \psi(\bB\btheta^{i})),
\end{align*}
where $\bPhi = (\bW+ \lambda \bI)^{-1} \bW$.
From \eqref{eq:k_bnd_nuproof3_1}, one can show that
\[
\bPhi = \bV_1 \tilde{\bD} \bV_1^\top,
\]
where $\tilde{\bD} = \Diag(\alpha_1^{-1},\ldots,\alpha_r^{-1})$,  $\alpha_i = 1 + \lambda \sigma_i^{-2}$; and then
\begin{align*}
\bV_1^\top \bPhi^i & = \bV_1^\top (\bV_1 \tilde{\bD}^i \bV_1^\top) = \tilde{\bD}^i \bV_1^\top
\end{align*}
The above results lead to
\ifconfver
\begin{align}
& \bV_1^\top (\btheta^k - \btheta^0 ) \nonumber \\
& ~~ = ( \tilde{\bD}^k - \bI ) \bV_1^\top \btheta^0 + \textstyle \sum_{i=0}^{k-1} \tilde{\bD}^{k-i} \bV_1^\top \bW^\dag \bB^\top \psi(\bB\btheta^{i}). \nonumber
\end{align}
\else
\begin{align}
	 \bV_1^\top (\btheta^k - \btheta^0 ) = (\tilde{\bD}^k - \bI ) \bV_1^\top \btheta^0 + \textstyle \sum_{i=0}^{k-1} \tilde{\bD}^{k-i} \bV_1^\top \bW^\dag \bB^\top \psi(\bB\btheta^{i}). \nonumber
\end{align}
\fi
Subsequently we have
\ifconfver
\begin{align}
	&\| \bV_1^\top (\btheta^k - \btheta^0 ) \|
 \nonumber \\
	& ~ \leq ( 1 - \alpha_r^{-k} ) \| \bV_1^\top \btheta^0 \|
	+ \textstyle \sum_{i=0}^{k-1} \alpha_1^{-(k-i)} \| \bV_1^\top \bW^\dag \bB^\top \psi(\bB\btheta^{i}) \| \nonumber \\
	& ~ \leq ( 1 - \alpha_r^{-k} ) \| \bV_1^\top \btheta^0 \|
	+ \textstyle \sum_{i=0}^{k-1} \alpha_1^{-(k-i)} \sigma_r^{-1} \sqrt{m} \nonumber \\
	& ~  = ( 1 - \alpha_r^{-k} ) \underbrace{\| \bV_1^\top \btheta^0 \|}_{:=a_1} + (1 - \alpha_1^{-k} ) \underbrace{\sigma_1^2 \lambda^{-1} \sigma_r^{-1} \sqrt{m}}_{:= a_2},
	\label{eq:k_bnd_nuproof4_1}
\end{align}
\else
\begin{align}
	\| \bV_1^\top (\btheta^k - \btheta^0 ) \|	
	&  \leq ( 1 - \alpha_r^{-k} ) \| \bV_1^\top \btheta^0 \|
	+ \textstyle \sum_{i=0}^{k-1} \alpha_1^{-(k-i)} \| \bV_1^\top \bW^\dag \bB^\top \psi(\bB\btheta^{i}) \| \nonumber \\
	&  \leq ( 1 - \alpha_r^{-k} ) \| \bV_1^\top \btheta^0 \|
	+ \textstyle \sum_{i=0}^{k-1} \alpha_1^{-(k-i)} \sigma_r^{-1} \sqrt{m} \nonumber \\
	&   = ( 1 - \alpha_r^{-k} ) \underbrace{\| \bV_1^\top \btheta^0 \|}_{:=a_1} + (1 - \alpha_1^{-k} ) \underbrace{\sigma_1^2 \lambda^{-1} \sigma_r^{-1} \sqrt{m}}_{:= a_2},
	\label{eq:k_bnd_nuproof4_1}
\end{align}
\fi
where the second inequality is due to $\bV_1^\top \bW^\dag \bB^\top = \tilde{\bSig}^{-1} \bU_1^\top$, which can be verified from \eqref{eq:k_bnd_nuproof3_1},
 and $\| \psi(\bB\btheta) \| \leq \sqrt{m}$.

Next, by applying \eqref{eq:k_bnd_nuproof4_1} to \eqref{eq:k_bnd_nuproof3_3a}, we obtain

\beq \label{eq:k_bnd_nuproof4_2}
\frac{( 1 - \alpha_r^{-k} ) a_1 + (1 - \alpha_1^{-k} ) a_2 }{a_1 + a_2}
\geq \frac{ \| \bV_1^\top(\btheta^\star - \btheta^0) \| - \eps }{a_1 + a_2} := C.
\eeq
Assume $C < 1$. Then,
from \eqref{eq:k_bnd_nuproof4_2},
\begin{align}
	\log ( 1 - C ) & \geq \log \left( \frac{ \alpha_r^{-k}  a_1 + \alpha_1^{-k} a_2  }{a_1 + a_2}  \right)
	\nonumber \\
	&
	\geq \frac{ a_1 \log( \alpha_r^{-k} ) + a_2 \log( \alpha_1^{-k} ) }{ a_1 + a_2},
	\label{eq:k_bnd_nuproof4_3}
\end{align}
where we use Jensen's inequality to obtain the second inequality.
Applying $\log(z) \leq z - 1$ to both sides of \eqref{eq:k_bnd_nuproof4_3} gives
\begin{align*}
& -C \geq -k \frac{ a_1 (\alpha_r-1) + a_2 (\alpha_1-1)}{ a_1 + a_2 } =
-k \frac{ a_1 \lambda \sigma_r^{-2} + a_2 \lambda \sigma_1^{-2}}{ a_1 + a_2 }
\\
& \Longrightarrow -(\| \bV_1^\top(\btheta^\star - \btheta^0) \| - \eps) \geq - k (\lambda \sigma_r^{-2}  \| \bV_1^\top\btheta^0 \| + \sigma_r^{-1} \sqrt{m} ) \\
& \Longrightarrow k \geq \sigma_r^2  \frac{  \| \bV_1^\top(\btheta^\star - \btheta^0) \| - \eps }{ \lambda \| \bV_1^\top\btheta^0 \| + \sigma_r \sqrt{m} },
\end{align*}
which is the desired result.
We also need to show $C < 1$.
As a variant of \eqref{eq:k_bnd_nuproof2_8}, we have
\begin{align*}
\| \bV_1^\top(\btheta^\star - \btheta^0) \|
& \leq \| \bV_1^\top \btheta^0 \| + \|  \btheta^\star \|
\leq \| \bV_1^\top \btheta^0 \| + \sigma_1 \sqrt{m}  \lambda^{-1} \\
& \leq \| \bV_1^\top \btheta^0 \| + \sigma_1 (\sigma_1/\sigma_r) \sqrt{m}  \lambda^{-1}  = a_1 + a_2.
\end{align*}
It follows that the $C$ in \eqref{eq:k_bnd_nuproof4_2} satisfies $C < 1$ for $\eps > 0$. The proof is complete.

\bibliographystyle{IEEEtran}
 \bibliography{ref}

\clearpage
\setcounter{page}{1}
\title{Supplemental Material of {``Accelerated and Deep Expectation Maximization for One-Bit MIMO-OFDM Detection''}}
\maketitle
\renewcommand\thesection{\arabic{section}}
\setcounter{section}{0}

\section{Computation of the Gradient in \eqref{eq:exa_grad}}

We describe how the gradient $\nabla f(\btheta)$ in \eqref{eq:exa_grad} can be efficiently computed.
Recall $\btheta= (\Re(\bs),\Im(\bs))$, and let $f(\bs)= f(\btheta)$ for convenience.
Let $\nabla_{\bs_n} f(\bs) = \nabla_{\Re(\bs_n)} f(\bs) + \jj \cdot \nabla_{\Im(\bs_n)} f(\bs)$ denote the gradient of $f$ w.r.t. $\bs_n$,
where $\nabla_{\Re(\bs_n)} f(\bs)$ and $\nabla_{\Im(\bs_n)} f(\bs)$ denote the gradient w.r.t. the real and imaginary parts of $\bs_n$, respectively.
Working through the details (see \eqref{eq:exa_grad}, \eqref{eq:model_uq}, \eqref{eq:model_uq_2}, and \eqref{eq:model_q_1}), one can show that
\begin{subequations} \label{eq:grad_comp}
	\begin{align}
		\nabla_{\bs_n} f(\bs) & = \textstyle -\sum_{m=1}^M \bF \bH_{m,n}^\her \bzeta_m^k,
		\label{eq:grad_comp:a} \\
		\Re( \bzeta_m^k ) & = \tfrac{1}{\sigma} \left( \Re(\bq_m) \odot \psi\left( \tfrac{1}{\sigma} \Re(\bq_m) \odot \Re(\bz_m^k) \right)  \right),
		\label{eq:grad_comp:c}
		\\
		\Im( \bzeta_m^k ) & = \tfrac{1}{\sigma} \left( \Im(\bq_m) \odot \psi\left( \tfrac{1}{\sigma} \Im(\bq_m) \odot \Im(\bz_m^k) \right)  \right),
		\label{eq:grad_comp:c2}
		\\
		\bz_m^k & = \textstyle \sum_{n=1}^N \bH_{m,n}\bF^\her \bs_n,
		\label{eq:grad_comp:d}
	\end{align}
\end{subequations}
for all $m,n$, where $\psi$ is defined in \eqref{eq:psi_def}.
Eqs.~\eqref{eq:grad_comp:a}  and \eqref{eq:grad_comp:d} are more expensive  to compute, at first sight, but one can utilize the OFDM structure to compute them efficiently.
Using \eqref{eq:H_evd}, we can rewrite \eqref{eq:grad_comp:a} and \eqref{eq:grad_comp:d} as
\begin{align}
\nabla_{\bs_n} f(\bs) & = \textstyle  \sum_{m=1}^M \bD_{m,n}^\her (\bF \bzeta_m^k), \label{eq:grad_comp:a_alt} \\
\bz_m^k & = \textstyle   \bF^\her (\sum_{n=1}^N \bD_{m,n} \bs_n),
\label{eq:grad_comp:d_alt}
\end{align}
respectively.
We can use IFFT and FFT to implement \eqref{eq:grad_comp:a_alt}--\eqref{eq:grad_comp:d_alt}.
In summary, the gradient $\nabla f(\btheta)$ in \eqref{eq:exa_grad} can be 
computed
using
$2M$ FFTs/IFFTs; we also need numerical computations of $\Phi$ for $2MW$ times (see \eqref{eq:grad_comp:c}--\eqref{eq:grad_comp:c2}).


\section{A Discussion on the Applications of the Existing Convergence Analysis Results}


In this section we examine and discuss the applications of the existing convergence analysis results to the EM method, with an emphasis on our problem.
The EM convergence was first studied by Wu \cite{wu1983convergence} in the statistics field.
EM is a special case of the MM method,
and there is a rich collection of literature on the convergence of the MM and related first-order methods in the optimization field; see, e.g., \cite{razaviyayn2013unified,mairal2013optimization,hong2017iteration} for studies closer to the context of EM.
In these MM studies we often consider a general framework that covers a variety of optimization schemes and a rich class of problems.
In that regard, we should point out that we focus on a specific problem arising from MIMO detection or regression from quantized data, and exploit opportunities unique to our problem.

Let us be specific.
The convergence analyses by Wu \cite{wu1983convergence} and  Razaviyayn {\em et al.} \cite{razaviyayn2013unified} show that the EM or MM method can converge to a stationary point. They consider the non-convex case with fairly general assumptions,
but the convergence rate is not considered.
The MM convergence rate in the convex case was studied by Mairal \cite{mairal2013optimization} and Hong {\em et al.} \cite{hong2017iteration}.
If we apply the analysis framework by Hong {\em et al.} to the EM method \eqref{eq:ana_EM} for our problem \eqref{eq:ana_P}, we have the following result:
\beq \label{eq:con_MM}
F(\btheta^k) - F(\btheta^\star) \leq \frac{\sigma_{\rm max}(\bB)^2}{k}  
\kappa(\bB)^2
C,
\eeq
for some constant $C \geq 0$,
where $\kappa(\bB)= \sigma_{\rm max}(\bB)/\sigma_{\rm min}(\bB)$ is the condition number of $\bB$.
Also, the application of Mairal's framework  leads to
	\beq \label{eq:con_MM2_t}
F(\btheta^k) - F(\btheta^\star) \leq \frac{\sigma_{\rm max}(\bB)^2}{k+1}
(2R^2),
\eeq
for some constant $R > 0$.
We will give the details of \eqref{eq:con_MM}--\eqref{eq:con_MM2_t} shortly.
The result in \eqref{eq:con_MM} is weak compared to our EM result in \eqref{eq:ana_EM_con};
in fact, it is even weaker than the PG result in \eqref{eq:ana_PG_con_alt}.
Also, as discussed in the main manuscript, the result in \eqref{eq:con_MM2_t} is arguably not as good as our EM result in \eqref{eq:ana_EM_con}.
\vspace{1em}

\subsection{Proof of \eqref{eq:con_MM}}

Consider problem \eqref{eq:ana_P} and the accompanied problem setup, and consider
the EM method \eqref{eq:ana_EM}.
Assume the following:
\begin{enumerate}[(a)]
	\item $\{ \btheta \in \Rbb^n | F(\btheta) \leq F(\btheta^0) \}$ is compact with diameter $R$;
	\item $g(\btheta|\btheta')$ is $\gamma$-strongly convex w.r.t. $\btheta$, i.e.,
	\[
	g(\btheta|\btheta') \geq g(\bvartheta|\btheta') + \langle \nabla g(\bvartheta|\btheta'), \btheta - \bvartheta \rangle + \tfrac{\gamma}{2} \| \btheta - \bvartheta \|^2,
	\]
	for all $\btheta, \bvartheta \in \Rbb^n$ and for any given $\btheta'$;
	\item $\nabla g(\btheta|\btheta')$ is $\beta$-Lipschitz continuous w.r.t. $\btheta' \in \Rbb^n$, i.e.,
	\[
	\| \nabla g(\btheta|\btheta') - \nabla g(\btheta|\btheta'') \| \leq \beta \| \btheta' - \btheta'' \|,
	\]
	for all $\btheta', \btheta'' \in \Rbb^n$ and for any given $\btheta$.
\end{enumerate}
The framework by Hong {\em et al.} \cite{hong2017iteration} gives the following result
\beq \label{eq:proof_MM_bnd}
F(\btheta^k) - F(\btheta^\star) \leq \frac{C_1}{\mu} \frac{1}{k}, \quad k \geq 1,
\eeq
where $C_1 = \max \{ 4\mu - 2, F(\btheta^0) - F(\btheta^\star), 2 \}$, and
\beq
\mu = \frac{\gamma}{2\beta^2 R^2}.
\nonumber
\eeq
The question is what are the values of $\gamma$ and $\beta$ for our problem.
It is seen from \eqref{eq:ana_u_EM} that $\gamma= \sigma_{\rm min}(\bB)^2$.
Also, it can be shown from \eqref{eq:lem:1:proof_dg} and  Proposition~\ref{prop:Lf} that $\beta =2  \sigma_{\rm max}(\bB)^2$.
Applying the above results to \eqref{eq:proof_MM_bnd} leads to \eqref{eq:con_MM}.

\subsection{Proof of \eqref{eq:con_MM2_t}}

Again, consider problem \eqref{eq:ana_P} and the accompanied problem setup, and consider
the EM method \eqref{eq:ana_EM}.
Let us assume:
\begin{enumerate}[(a)]
	\item $\{ \btheta \in \Rbb^n | F(\btheta) \leq F(\btheta^0) \}$ is compact with diameter $R$;
	\item $r(\btheta|\btheta') := g(\btheta|\btheta') - f(\btheta)$ has $L_r$-Lipschitz continuous gradient w.r.t $\btheta$ and for any given $\btheta'$.
\end{enumerate}
Mairal's framework \cite{mairal2013optimization} gives the following result
\beq
F(\btheta^k) - F(\btheta^\star) \leq  \frac{2 L_r R^2}{k+1}, \quad k \geq 1. \nonumber
\eeq
To show the value of $L_r$ for our problem, we first note that $f$ is convex and has $\sigma_{\rm max}(\bB)^2$-Lipschitz continuous gradient (cf. Proposition~\ref{prop:Lf} for the latter).
From the expression of $g$ in \eqref{eq:ana_u_EM}, one can easily verify that $g(\btheta|\btheta')$ is convex w.r.t. $\btheta$ and has $\sigma_{\rm max}(\bB)^2$-Lipschitz continuous gradient w.r.t. $\btheta$, given any $\btheta'$.
By Lemma B.9 in the supplementary material of \cite{mairal2013optimization},
$r(\btheta|\btheta')$ has $\sigma_{\rm max}(\bB)^2$-Lipschitz continuous gradient w.r.t. $\btheta$.
We therefore have \eqref{eq:con_MM2_t}.

\section{Proof of \eqref{eq:check_err}}

Let $\chi(g,\theta)= \min_{v \in \partial h(\theta) } | g + v |$,
$h(\theta) = \Ind_{[-U,U]}(\theta)$. Since
\beq \label{eq:err_proof1} \nonumber
\partial h(\theta) = \left\{  \begin{array}{ll}
	\Rbb_+, & \theta = U \\
	\Rbb_-, & \theta = -U \\
	\{ 0 \}, & |\theta| < U
\end{array} \right.
\eeq
one can verify that
\beq\label{eq:err_proof2} \nonumber
\chi(g,\theta) = \left\{  \begin{array}{ll}
	|g|, & \text{$g \geq 0$ and $\theta = U$, or $g \leq 0$ and $\theta = -U$} \\
	0,   & \text{$g < 0$ and $\theta = U$, or $g > 0$ and $\theta = -U$} \\
	|g|, & |\theta| < U
\end{array} \right.
\eeq
Applying the above result to \eqref{eq:aiem_sol_req2}, with the separable property $h(\btheta) = \sum_{i=1}^n h(\theta_i)$ taking into consideration, we obtain \eqref{eq:check_err} as the equivalent form of \eqref{eq:aiem_sol_req2}.

\section{Additional Numerical Results}

In this section we provide additional numerical results to illustrate the behaviors of the PG and EM methods.


	\subsection{Convergence Behaviors}
	
\begin{figure}[h]
		\centering
		\includegraphics[width=0.7\linewidth]{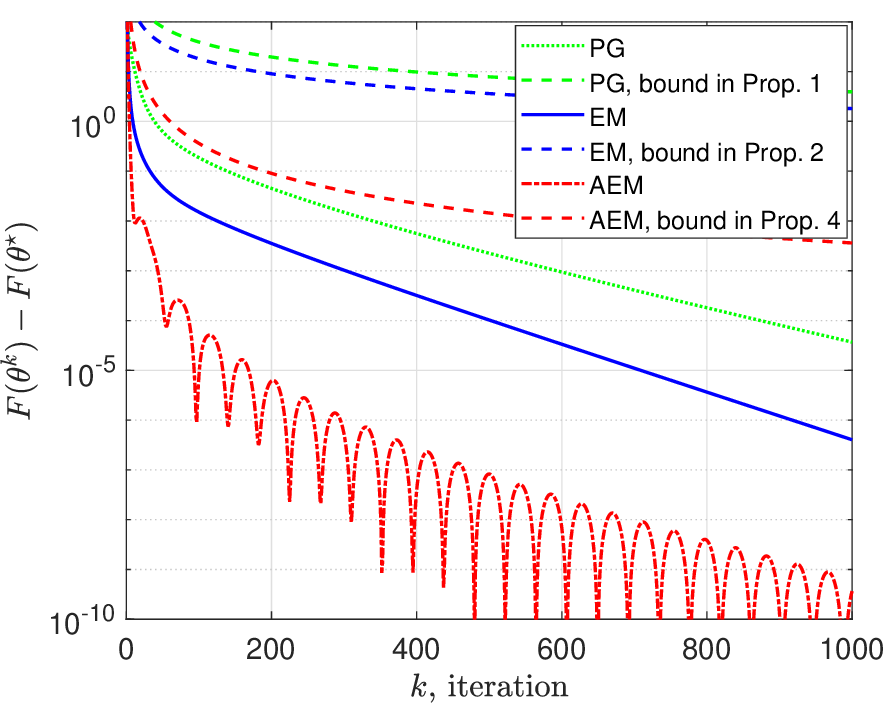}
		\caption{Objective value gaps of PG, EM and AEM. }\label{fig:conv_rate}
	\end{figure}

	We want to see whether the PG, EM and AEM methods behave as predicted by the theoretical convergence results in Propositions \ref{prop:Lf}, \ref{prop:EM}, \ref{prop:new_EM_PG} and \ref{prop:AEM}.
	We first consider the theoretical upper bound results in Propositions \ref{prop:Lf}, \ref{prop:EM}, and \ref{prop:AEM}.
	To do this, we randomly generated an instance $(\by,\bA,\btheta)$ in accordance with the basic QLR model in \eqref{eq:exa_1bit_model}, with $(m,n)= (16,4)$ and with the elements of $\btheta$ and $\bA$ being i.i.d. $\setN(0,1)$.
	Then we ran the PG, EM and AEM methods for problem  \eqref{eq:ana_P} with the penalty function being $h(\btheta) = \| \btheta \|^2/2$ and with the initialization being randomly generated.
	Fig.~\ref{fig:conv_rate} shows the objective value gaps $F(\bm \theta^k)-F(\bm \theta^\star)$ of the three methods and their respective theoretical upper bounds in Propositions \ref{prop:Lf}, \ref{prop:EM},  and \ref{prop:AEM};
	note that this is a one-trial result.
	We see that the PG, EM and AEM methods appear to converge at rates that are at least no worse than those of the respective theoretical upper bounds, which gives support to our theory.
	We also notice that the objective value gaps of the three methods are far better than the respective theoretical upper bounds, suggesting that the three methods may converge faster than what theory predicts.
	
\begin{figure}[h]
		\centering
		\includegraphics[width=0.7\linewidth]{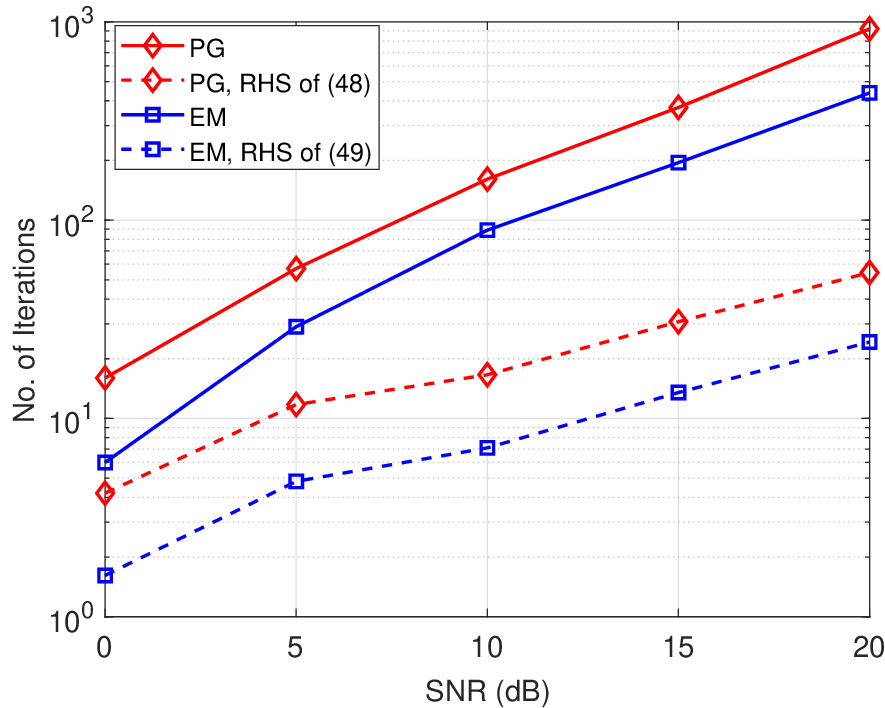}
		\caption{Iteration numbers of PG and EM. }\label{fig:conv_LB}
	\end{figure}

	Next we consider the theoretical lower bound results in Proposition  \ref{prop:new_EM_PG}.
	We use the same method as above to perform numerical experiments.
	For each method, we evaluate the smallest number of iterations required to achieve $\|\bm \theta^k - \bm \theta^{\star} \|\leq \epsilon$, with $\epsilon = 10^{-2}$.
	Fig.~\ref{fig:conv_LB} shows the iteration numbers of the PG and EM methods and their respective theoretical lower bounds in Proposition  \ref{prop:new_EM_PG}; again it is a one-trial result.
	We see that the iteration numbers of the PG and EM methods increase with the SNR,
	and the rates of increase appear to be consistent with those of the theoretical lower bounds.

%
	
\subsection{Impact of Noise Power Loading}

\begin{figure}[htb!]
\centering
\includegraphics[width=0.7\linewidth]{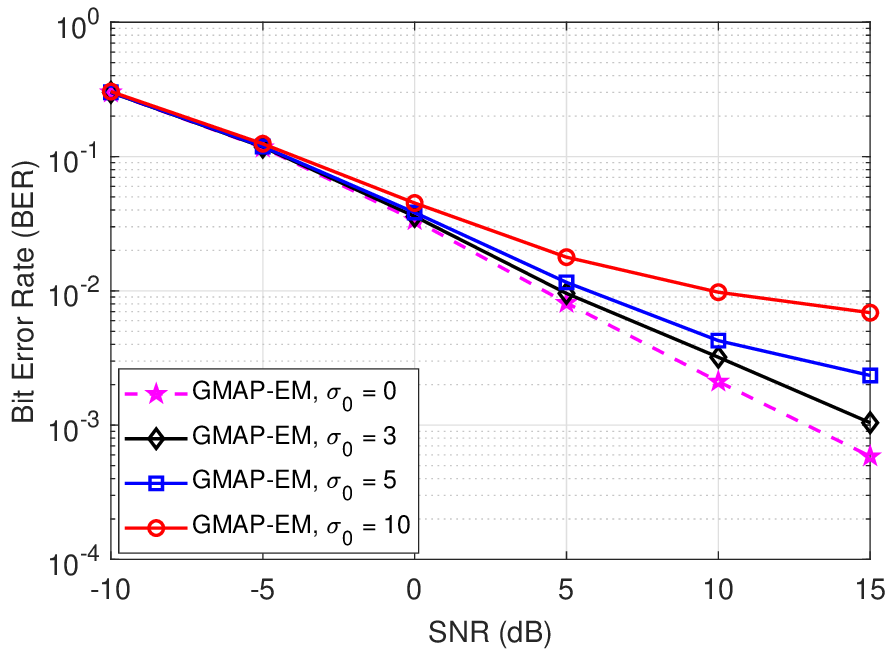}
 \caption{BER performance of GMAP-EM for various $
 \sigma_0$.  $(M,N,W)=(128,10,256)$, $16$-QAM.}\label{fig:sigma_plot}
\end{figure}

We examine the impact of noise power loading in \eqref{eq:loading}.
The simulation settings are identical to those in Section~\ref{sec:sim}.
For brevity we consider GMAP-EM only; the other algorithms exhibit similar behaviors, as we observed by numerical study.
 Fig.~\ref{fig:sigma_plot} shows the BER performance for various values of the loading parameter $\sigma_0$.
We see performance loss when loading is applied; the loss increases with $\sigma_0$.
Table~\ref{tb:iter_sigma} shows the average number of iterations for various values of $\sigma_0$ and the SNR.
We see that, without noise power loading, convergence is particularly slow for high SNRs;
and that such undesirable effects are alleviated when loading is applied.

\begin{table}[htb!]
\centering
\caption{Average EM iterations of GMAP-EM under different $\sigma_0$. $(M,N,W)=(128,10,256)$, $16$-QAM. }\label{tb:iter_sigma}
\resizebox{0.8\linewidth}{!}{
\begin{tabular}{M{38mm}|M{8mm}|M{8mm}| M{8mm} |M{8mm}|M{8mm}|M{8mm}}
\hline  
  \backslashbox{$\sigma_0$}{SNR (dB)}   &   -10      &   -5    & 0 & 5 & 10 & 15 \\
\hline \hline 
$\sigma_0 = 0$& 22.7 & 	55.9 &	140 &	292.1 & 476.1 &	594.6  \\
\hline 
$\sigma_0 = 3$ & 20.9 &	46.7 &	97.5 & 165.4 &	240 &	296 \\
\hline 
$\sigma_0 = 5$ &19.9 &	41.8 &	79.6 &	123 &	167.5 &	199 \\
\hline 

$\sigma_0 = 10$ &17.7 &	32.6 &	52.5 & 70.7 &  87.5 & 97.5 \\
\hline 

\end{tabular}
}
\end{table}


\end{document}